\documentclass[journal]{IEEEtran}
\usepackage[dvipsnames]{xcolor}
\definecolor{c1}{HTML}{177cb0}
\usepackage{url}
\usepackage{tabularx}
\usepackage{float}
\usepackage{graphicx}
\usepackage{nomencl}
\usepackage{amsmath}
\usepackage{booktabs}
\usepackage{stfloats}
\usepackage{amssymb}
\usepackage{multirow}
\usepackage{subfigure}
\usepackage{xcolor}
\usepackage{threeparttable}
\usepackage{utfsym}
\usepackage{wasysym}
\usepackage{algorithm}
\usepackage{algpseudocode}
\usepackage{ulem}

\usepackage{cite}

\ifCLASSINFOpdf
\else
\fi
\begin{document}
%
\title{Reliability-Based Planning of Cable Layout for Offshore Wind Farm Electrical Collector System Considering Post-Fault Network Reconfiguration}
%
%
%

\author{Xiaochi~Ding,
Yunfei~Du,
Xinwei~Shen,~\IEEEmembership{Senior~Member,~IEEE,}
Qiuwei~Wu,~\IEEEmembership{Senior~Member,~IEEE,}
Xuan~Zhang,~\IEEEmembership{Senior~Member,~IEEE,}
Nikos D.~Hatziargyriou,~\IEEEmembership{Life~Fellow,~IEEE}
\vspace{-1em}

\thanks{Xiaochi Ding, Qiuwei Wu, and Xuan Zhang are with Tsinghua-Berkeley Shenzhen Institute, Tsinghua Shenzhen International Graduate School (SIGS), Tsinghua University. Yunfei Du and Xinwei Shen are with Institute for Ocean Engineering, Tsinghua SIGS, Tsinghua University. N. D. Hatziargyriou is with the School of Electrical and Computer Engineering, National Technical University of Athens. This work is supported by National Natural Science Foundation of China (52477102), Excellent Youth Basic Research Fund of Shenzhen (RCYX20231211090430053) and Guangdong Basic and Applied Basic Research Foundation (2022A1515240019, 2023A1515240055). (Corresponding author: Xinwei Shen, xwshen@tsinghua.edu.cn).
}%

}

%
%

\markboth{Journal of \LaTeX\ Class Files,~Vol.~14, No.~8, August~2015}%
{Shell \MakeLowercase{\textit{et al.}}: Bare Demo of IEEEtran.cls for IEEE Journals}
%



\maketitle

\begin{abstract}
The electrical collector system (ECS) plays a crucial role in determining the performance of offshore wind farms (OWFs). 
Existing research has predominantly restricted ECS cable layouts to conventional radial or ring structures and employed graph theory heuristics for solutions.
However, both economic efficiency and reliability of the OWFs heavily depend on their ECS structure, and the optimal ECS cable layout often deviates from typical configurations.
In this context, this paper introduces a novel reliability-based ECS cable layout planning method for large-scale OWFs, employing a two-stage stochastic programming approach to address uncertainties of wind power and contingencies. To enhance reliability, the model incorporates optimal post-fault network reconfiguration strategies by adjusting wind turbine power supply paths through link cables. To tackle computation challenges arising from numerous contingency scenarios, a customized progressive contingency incorporation (CPCI) framework is developed to solve the model with higher efficiency by iteratively identifying non-trivial scenarios and solving the simplified problems. The convergence and optimality are theoretically proven.
Numerical tests on several real-world OWFs validate the necessity of fully optimizing ECS structures and demonstrate the efficiency of the CPCI algorithm.

\end{abstract}

\vspace{-2em}
\section*{Nomenclature} 
\subsection*{Indices and Sets} 
\addcontentsline{toc}{section}{Nomenclature}
\begin{IEEEdescription}[\IEEEusemathlabelsep\IEEEsetlabelwidth{$i, j, x, y$}] 
\item[$br^f$] Index for the cable connected to feeder $f$.
\item[$f$] Index for main feeder.
\item[$i, j, r, s, k$] Indices for nodes.
\item[$ij, rs$] Indices for cables.
\item[$NO$] Index for normal operational state.
\item[$\upsilon/\omega$] Index for operational state / wind speed scenario.
\item[$\Psi_N/\Psi_N^{WT}$] Set of nodes / wind turbine nodes.
\item[$\Psi_C/\Psi_F$] Set of candidate cables / main feeders.
\item[$\Psi_i$] Set of nodes connected to node $i$.
\item[$\Upsilon$] Set of system operational state scenarios.
\item[$\Omega$] Set of wind speed scenarios.
\item[$\mathcal{X}_C$] Set of crossing cable pairs.
\vspace{-1.5em}
\end{IEEEdescription}
\subsection*{Parameters} 
\addcontentsline{toc}{section}{Nomenclature}
\begin{IEEEdescription}[\IEEEusemathlabelsep\IEEEsetlabelwidth{$i, j, x, y$}] 
\item[$C_{cab}$] Cost of purchasing and installing one unit length of submarine cable with switches.
\item[$C_{ele}$] Unit price of offshore wind energy.
\item[$a_{ij}$] Length of candidate cable $ij$.
\item[$\beta$] Ratio of $C_{O\&M}$ to $C_{INV}$.
\item[$H$] Hours in one year, i.e., 8760.
\item[$\delta^\omega$] Probability of wind speed scenario $\omega$.
\item[$\zeta^\omega$] Magnitude of wind speed scenario $\omega$.
\item[$\xi^{rs/NO}$] Probability of cable $rs$ fault / $NO$ scenario.
\item[$q/t$] Discount ratio / Operation time of the project.
\item[$B_{ij}$] Susceptance of cable $ij$.
\item[$M$] Big-M constant.
\item[$P_f^C/P_{ij}^C$] Power transmission capacity of feeder $f$/cable $ij$.
\item[$P_k$] Rated capacity of wind turbine $k$.
\item[$\lambda_{ij} / \mu_{ij}$] Failure / Repair rate of cable $ij$.
\item[$\tau^{SW}/\tau^{RP}$] Time required to isolate / repair the cable fault.
\end{IEEEdescription}
\vspace{-1.5em}
\subsection*{Binary Variables} 
\addcontentsline{toc}{section}{Nomenclature}
\begin{IEEEdescription}[\IEEEusemathlabelsep\IEEEsetlabelwidth{$i, j, x, y$}]
\item[$l_{ij}$] Cable investment variable, 1 when cable $ij$ is invested for the installation.
\item[$m_k^{NO/rs}$] Fault impact variable, 1 when wind turbine $k$ is affected in $NO$ / cable $rs$ contingency scenario.
\item[$n_k^{NO/rs}$] Fault continuation variable, 1 when wind turbine $k$ cannot send power in $NO$ / after network reconfiguration following cable $rs$ fault.
\item[$s_{ij}^{NO/rs}$] Connection status of cable $ij$ in $NO$ / after reconfiguration following cable $rs$ fault.
\item[$h_{ij}^f/h_k^f$] Cable-feeder / Node-feeder affiliation variable, 1 when cable $ij$ / wind turbine $k$ supplies power to the substation through feeder $f$ in $NO$.
\end{IEEEdescription} 
\vspace{-1.5em}
\subsection*{Continuous Variables} 
\addcontentsline{toc}{section}{Nomenclature}
\begin{IEEEdescription}[\IEEEusemathlabelsep\IEEEsetlabelwidth{$i, j, x, y$}] 
\item[$C_{INV}$] Investment cost for constructing collector system.
\item[$C_{O\&M}$] Operation and maintenance cost of the system.
\item[$C_{REL}$] Reliability cost resulting from wind energy curtailment caused by cable contingencies.
\item[$P_f^{rs}/P_{ij}^{rs}$] Power flowing through feeder $f$ / cable $ij$ after reconfiguration due to fault of cable $rs$.
\item[$P_k^{rs}$] Wind power sent by wind turbine $k$ after reconfiguration following fault of cable $rs$.
\item[$\theta_i^{rs}$] Voltage phase of $i$ when cable $rs$ fails.
\end{IEEEdescription} 

%
\IEEEpeerreviewmaketitle


\section{Introduction}



\IEEEPARstart{E}{nhancing} the utilization of offshore wind resources is crucial for achieving the global energy transition to fight climate change. 
Currently, offshore wind farms (OWFs) are expanding to more distant locations to harness more powerful wind resources, resulting in rising costs. 
In most OWF projects, the expenses related to wind turbines (WTs) account for 40-50\% of the total electrical infrastructure expenditure. And the costs of the electrical collector system (ECS), consisting of interconnecting cables between WTs and offshore substations (OSSs), account for 15-30\%\cite{zuo2022review}. 
However, the WT cost remains relatively fixed once their types and capacities are determined. In contrast, the flexible design of ECS layout could save substantial expenses, emphasizing the importance of its planning method. \par
The ECS serves to transmit the electricity generated by WTs to OSSs. The planning of ECS cable layout is characterized as NP-hard and non-convex\cite{zuo2022review,perez2019electrical}. As the scale of ECS increases, the complexity arises from the numerous feasible layout options available, making the selection of the optimal solution extremely challenging\cite{dutta2012optimal}. According to Ref \cite{jenkins2013offshore}, in a case involving 75 WTs, there are approximately $1 \times 10^{107}$ potential designs. It would take around $9 \times 10^{89}$ years to find the optimal solution by enumerating each feasible solution. This is clearly impractical, hence it is necessary to investigate optimization methods to obtain high-quality ECS layouts within an acceptable timeframe.\par
Moreover, the failure rates of submarine cables are significantly higher compared to their onshore counterparts due to the harsh sea conditions. The remote location and limited accessibility in adverse weather conditions result in longer repair time for faults, typically two to three months \cite{wei2016hierarchical}. The curtailment of wind energy during faults can be substantial and lead to severe economic losses. Thus, the reliability must be considered in the ECS planning. \par

There is a solid foundation in the research on ECS planning. Ref \cite{quinonez2007electrical} classifies typical ECS structures into four categories: string design, star design, single-sided ring design, and double-sided ring design. The radial structure requires lower investment, while the ring structure achieves higher reliability by allowing bidirectional power flow through backup cables.
Early research focused on the planning of radial ECS\cite{gonzalez2011optimal,dutta2012optimal,lumbreras2012optimal,lumbreras2012progressive,chen2016collector,cerveira2016optimal,shin2016optimal,fischetti2018optimizing,taylor2023wind,perez2019global,wang2023fast,song2023optimization,wei2023electrical,chen2024grouping,jin2019cable}.
With the development of long-distance large-scale OWFs, in which the reliability requirement is very high, researchers have gradually turned their attention to the ring ECS\cite{zuo2019collector,zuo2020collector,zuo2021two,dahmani2016optimization,wei2016hierarchical,perez2020closed,shen2023optimal}. One practical application is the London Array OWF.
However, while these studies of typical structures serve as valuable guidelines, it is worth noting that both economy and reliability are so dependent on ECS topology that the optimal structure seldom corresponds to the typical ones\cite{lumbreras2012optimal,gong2017optimal}. \par
To our knowledge, although ECSs with non-typical structures have been applied in practice (e.g., Triton Knoll OWF\cite{tritonknoll}), there is limited literature \cite{zuo2019collector,zuo2020collector,gong2017optimal} studying the benefits of non-typical structures from a theoretical view.
Refs \cite{zuo2019collector,zuo2020collector} innovatively introduced the Cross-Substation Incorporation (CSI) structure, highlighting its economic benefits in capacity sharing among OSSs and leveraging the reliability advantages of ring structure. 
The multi-loop structure was proposed in \cite{gong2017optimal} and demonstrated to be more economical than both string and ring structures when the cable failure rate and repair time are high. 
But these studies were still conducted within the predefined framework of CSI or multi-loop structures, and heuristic methods were adopted during the planning process. 
Hence, it is worthwhile to study how to optimize ECS cable layout by overcoming structural limitations to realize its ``free evolution" under various conditions, aiming to achieve the optimal balance between economy and reliability. \par
The majority of studies initially limit the scope of ECS planning to either radial or ring structure and formulate the problem as classical \textit{graph theory} problems, including Capacitated Minimum Spanning Tree (CMST)\cite{dutta2012optimal,shin2016optimal,zuo2020collector,chen2024grouping}, Multiple Traveling Salesman Problem (MTSP)\cite{gonzalez2011optimal,wei2016hierarchical}, or Capacitated Vehicle Routing Problem (CVRP)\cite{gong2017optimal,zuo2021two,shen2023optimal}. Once the problems are formulated, corresponding algorithms including the Prim, Clark and Wright's Savings (CWS), and genetic algorithms are applied to find solutions. The efficiency and ease of implementation of heuristic and metaheuristic methods make them more popular than mathematical programming (MP) ones in the existing literature\cite{perez2019electrical}. However, it is always hard to evaluate the quality of their solutions. Moreover, infeasible outputs may arise sometimes due to the challenges of implementing geographical restrictions \cite{zuo2022review}. 
In contrast, the MP method could avoid infeasible solutions by adding specific constraints and get optimal solutions. Hence, this paper adopts MP to optimize the ECS cable layout. \par
In addition, the trend in developing large-scale OWFs is to utilize multiple OSSs to collect the power from WTs. Only Refs \cite{gonzalez2011optimal,chen2016collector,dahmani2016optimization,perez2019global,zuo2019collector,wei2023electrical} have addressed the ECS design with multiple OSSs. 
Most studies initially cluster WTs, allocate them to different OSSs, and subsequently design the layouts independently. In other words, the interconnection between OSSs was overlooked.
The results in \cite{zuo2019collector} demonstrate that considering interconnections between OSSs facilitates OSS coordination and thereby enhances overall economic efficiency of ECSs. However, the adopted two-phase CWS algorithm is a heuristic approach, and as such, there is no guarantee of the planning results' optimality.\par

To address the aforementioned challenges encountered in state-of-the-art research, this paper proposes an innovative reliability-based planning method for ECS cable layouts. The contributions can be summarized as follows:
\begin{itemize}
    \item[1)]
    To the best of our knowledge, this work is the first attempt to realize ECS cable layout optimization without predefined radial or ring structural restrictions. 
    The full optimization of ECS expands the feasible region, which can strike a balance between economy and reliability more effectively than conventional planning approaches. 
    \item[2)] 
     A two-stage stochastic programming model is formulated to comprehensively address the uncertainties of wind speed and system faults, which considers the optimal post-fault network reconfiguration and coordination of OSSs to improve ECS economic efficiency and reliability.
    \item[3)] 
    A divide-and-conquer approach, including OWF partition and candidate cable generation strategies, is developed to perform dimension reduction. Moreover, a customized progressive contingency incorporation algorithm is devised to accelerate the solution process. The convergence and optimality of the algorithm are theoretically proven.
\end{itemize}
The remainder of the paper is organized as follows: Section II introduces the methodology framework and modeling details; Sections III and IV present the mathematical model and solution framework; case studies and discussions are given in Sections V and VI, and Section VII concludes the paper.
\section{Methodology Descriptions} 
\subsection{Methodology Framework}
\vspace{-1em}
\begin{figure}[!htbp]
\centering
\includegraphics[width=3.39in]{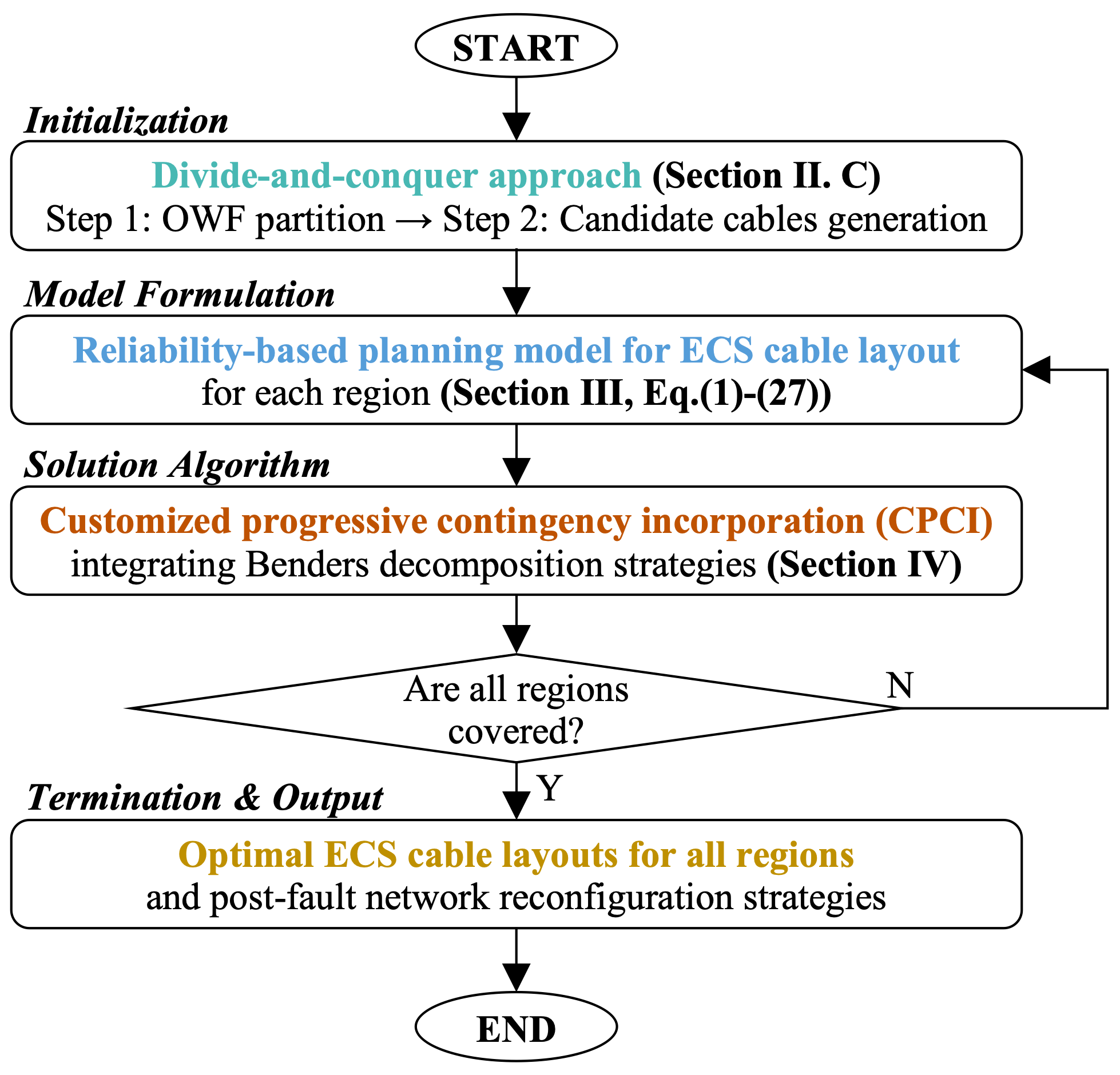}
\vspace{-1em}
\caption{Framework of the proposed ECS planning methodology. }
\label{Flowchart}
\vspace{-1em}
\end{figure}
The framework of the proposed methodology is shown in Fig. \ref{Flowchart}. In the first stage, the coordinates of WTs and OSSs are taken as inputs as they are determined in the micro-siting process. These coordinates are processed using the Divide-and-conquer approach to partition the OWF and generate candidate cables, thereby reducing the complexity. In the second stage, stochastic models are formulated to characterize the uncertainties associated with wind speeds and system faults. By exploiting the problem structure, a customized progressive contingency incorporation algorithm, embedded with Benders decomposition strategy, is applied to solve the large-scale MILP models in the third stage. Ultimately, the optimal cable layout and post-fault network reconfiguration strategies are obtained. Details of each stage will be elaborated in subsequent sections.


\vspace{-1em}

\subsection{ECS Post-Fault Network Reconfiguration Process}
\begin{figure}[!htbp]
\centering
\includegraphics[width=3.39in]{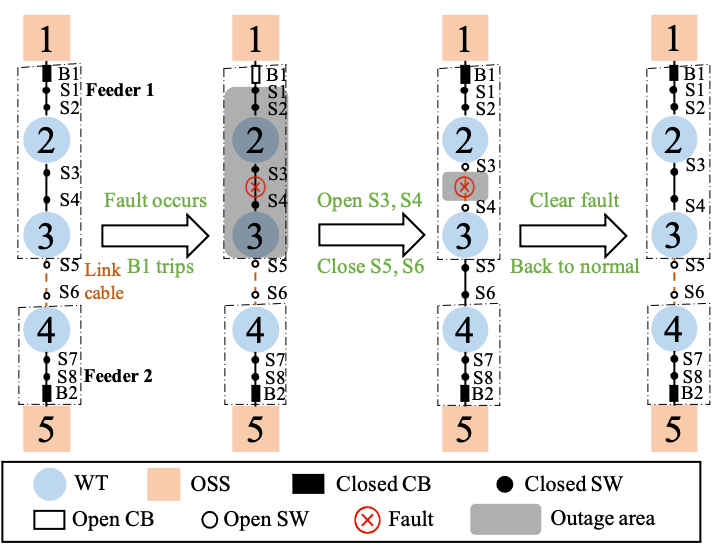}
\vspace{-0.8em}
\caption{Illustration of ECS post-fault network reconfiguration process.}
\vspace{-1em}
\label{5WTs}
\end{figure}

The ECS post-fault network reconfiguration process aligns with the approach outlined in \cite{ding2023smart}, as illustrated by a simple ECS in Fig. \ref{5WTs}. When a permanent fault occurs in cable 2-3, the circuit breaker (CB) B1 on Feeder 1 trips, causing a momentary power outage in Feeder 1. After a short period, the fault location is identified, and switches (SWs) S3 and S4 on both sides of the fault are opened to isolate the fault locally. S5 and S6 on the link cable 3-4 are closed to reconfigure the network and transfer WTs' power supply paths between feeders. Once the fault is cleared, the ECS restores its normal operation state.
Additional details can be found in \cite{ding2023smart}.
\vspace{-1em}

\subsection{Model Dimension Reduction}

\begin{figure*}[!htbp]
\vspace{-0.5em}
\centering
\includegraphics[width=0.7\textwidth]{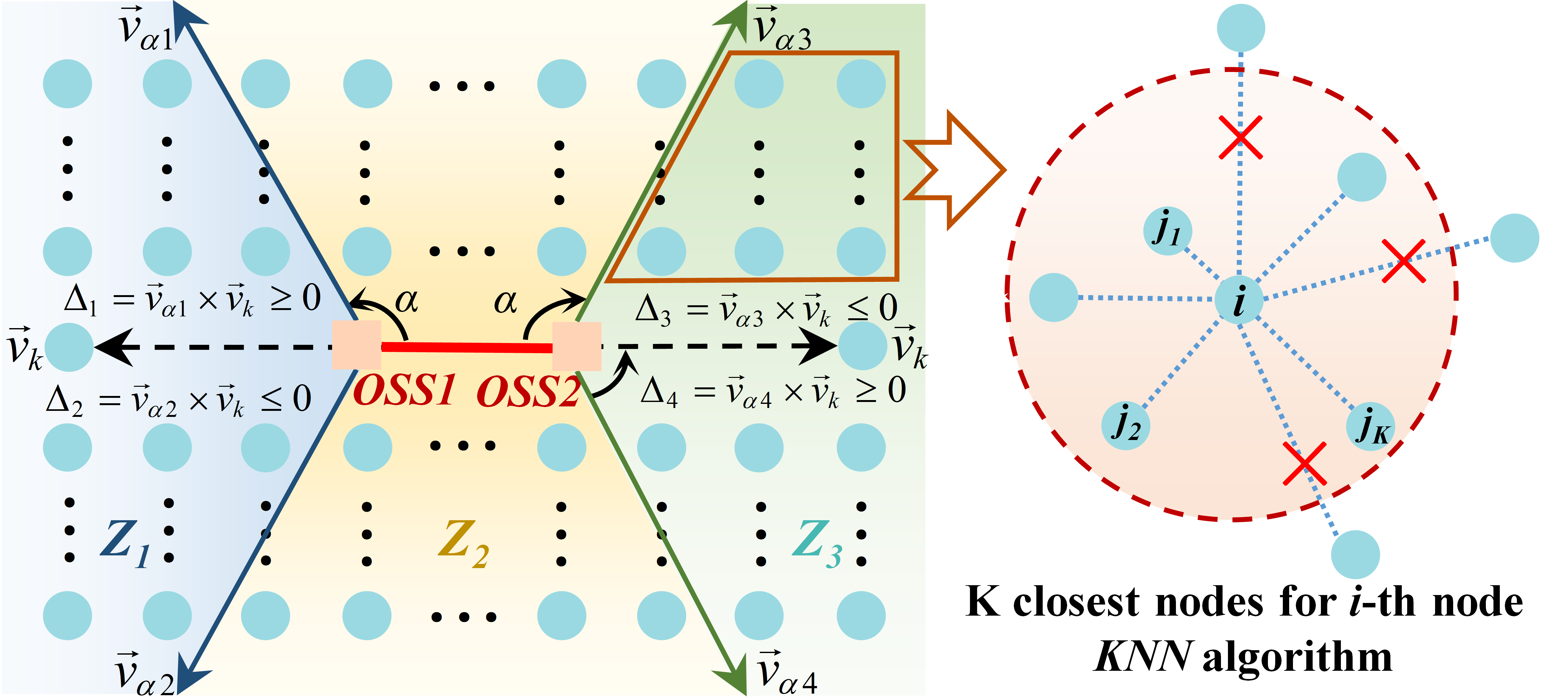}
\vspace{-1em}
\caption{Illustration of OWF partition and candidate cables generation.}
\label{DandC}
\vspace{-1em}
\end{figure*}

The ECS planning problem's complexity is associated with the number of candidate cables, namely, the dimension of decision variables in the optimization model. Obviously, the abundance of variables inevitably leads to long solution time.
Partitioning is a useful technique to simplify the modeling of large and intricate engineering systems, which has been implemented in various OWF planning problems\cite{zuo2019collector,zuo2021two,wang2023fast}. Therefore, a divide-and-conquer approach is developed here, as shown in Algorithm \ref{alg:DaC} and depicted in Fig. \ref{DandC}.\par
\floatname{algorithm}{Algorithm}
\renewcommand{\algorithmicrequire}{\textbf{Step 1: OWF partition}} 
\renewcommand{\algorithmicensure}{\textbf{Step 2: Candidate cables generation}}
\begin{algorithm}[htbp]\small   
\caption{ Divide-and-conquer Approach }    
\label{alg:DaC}                  
\begin{algorithmic}[1]                
\Require   
\State $\alpha=0$
\While {$\alpha \leq \bar\alpha$} 
\State $\mathcal{Z}_1=\mathcal{Z}_2=\mathcal{Z}_3=\varnothing$;
\For{$k = 1:\lvert \Psi_N^{WT} \rvert$}
    \If{${\triangle}_1=\Vec{v}_{a1}(\alpha) \times \Vec{v}_{k}\geq 0 \land {\triangle}_2=\Vec{v}_{a2}(\alpha) \times \Vec{v}_{k}\leq 0$} 
    \State $\mathcal{Z}_1=\mathcal{Z}_1 \cup k, t_{1k}=1$;
    \ElsIf {${\triangle}_3=\Vec{v}_{a3}(\alpha) \times \Vec{v}_{k}\leq 0 \land {\triangle}_4=\Vec{v}_{a4}(\alpha) \times \Vec{v}_{k}\geq 0$} 
    \State $\mathcal{Z}_3=\mathcal{Z}_3 \cup k, t_{3k}=1$;
    \Else
    \State $\mathcal{Z}_2=\mathcal{Z}_2 \cup k, t_{2k}=1$;
\EndIf
\EndFor
\State $U_{\alpha}=\max \left\{\left|\sum_{k=1}^{\lvert \Psi_N^{WT} \rvert} t_{i k}-\sum_{k=1}^{\lvert \Psi_N^{WT} \rvert} t_{jk} \right| \mid 1 \leq i, j \leq 3\right\}$;
\State $\alpha=\alpha+\Delta\alpha$;
\EndWhile
\State Partition OWF by optimal $\alpha$ with minimum $U_{\alpha}$;
\Ensure  
\State $\Psi_C=\varnothing$;
\For{$i = 1:\lvert \Psi_N \rvert$}
\State Apply \textit{KNN} algorithm to the $i$-th node and identify its K closest nodes $\{j_1,j_2,...,j_K\}$;
\State $\Psi_C=\Psi_C\cup\{ij_1,ij_2,...,ij_K\}$;
\EndFor
\end{algorithmic}
\end{algorithm}
\setlength{\intextsep}{3pt} 
In \textbf{Step 1} (lines 1-16), the entire OWF is divided into several computationally tractable regions, as illustrated by the left figure in Fig. \ref{DandC}. The $\alpha$ is defined as the angle between the boundary (denoted as the arrowed solid line) and the virtual connection between OSSs (denoted as the red line), and $\bar\alpha$ and $\Delta\alpha$ represent the upper limit and increment of $\alpha$. The algorithm first determines the affiliation relationship between WTs and Regions 1-3 ($\mathcal{Z}_1$-$\mathcal{Z}_3$) for different $\alpha$, and then calculates $U_{\alpha}$, representing the maximum difference in the number of WTs within each region. To achieve optimal division, the $\alpha$ corresponding to the minimum $U_\alpha$ is selected as the boundary for partitioning.
In \textbf{Step 2} (lines 17-21), candidate cables are generated, as shown in the right part of Fig. \ref{DandC}. The K-Nearst Neighbors (\textit{KNN}) algorithm is applied based on the plausible expectation that the optimal layout does not require long cables\cite{toth2003granular}.
Applying Algorithm \ref{alg:DaC} effectively reduces the dimensionality, leading to improved efficiency.
\section{Reliability-Based Planning Model \\ for ECS Cable Layout}
In this section, we develop a stochastic optimization model for ECS planning. The model could generate the optimal cable layout and reconfiguration strategies under contingencies. Due to the complexity, only one type of cable is temporarily considered, as done in\cite{taylor2023wind,shen2023optimal,wang2020optimization}. The notation of the symbols can be found in the Nomenclature.
\vspace{-1.5em}
\subsection{Model Assumptions}
The following assumptions are adopted for tractability. 
\begin{itemize}
\item[1)]
Given the voltage levels of the ECSs, the DC power flow model is adopted \cite{shen2023optimal}. The system is operated in a radial configuration\cite{gong2017optimal} to mitigate the risk of elevated fault currents that could arise during closed-loop operation \cite{lee2009a}.
\item[2)]
Given that ECS cables are buried beneath the seabed and consequently exhibit very low failure rates, the occurrence of simultaneous faults in multiple cables is highly unlikely. Therefore, the contingency set is limited to include only single cable outages, as reported in \cite{lumbreras2012progressive,shin2016optimal,shen2023optimal,gong2017optimal}.
\item[3)]
The set of various wind scenarios is produced by scenario generation and reduction techniques \cite{perez2019electrical,perez2020closed}.
\item[4)]
The discrete Markov Chain model is applied to calculate the probability of a cable being in the unavailable state, as per $\xi^{ij}=\frac{\lambda_{ij}}{\lambda_{ij}+\mu_{ij}}$\cite{calixto2016gas}, where $\lambda_{ij}, \mu_{ij}$ denote the failure rate and repair rate of cable $ij$. These rates are obtained from statistical data, as is common in most literature.

\end{itemize}
\vspace{-1.5em}
\subsection{Mathematical Formulation}

\begin{equation}
\min_{\left\{\substack{
l_{ij},s^{NO/rs}_{ij},m^{NO/rs}_{k},n^{NO/rs}_{k}
\\
h^{f}_{ij},h^{f}_{k},P^{rs}_{f},P^{rs}_{ij},P^{rs}_{k},\theta^{rs}_{i}
}\right\}} Obj=C_{INV}+C_{O\&M}+C_{REL} 
\end{equation}

\begin{equation}
C_{INV}=C_{c a b} \sum_{i j \in \Psi_C} a_{ij} l_{i j} 
\end{equation}
\begin{equation}
C_{O\&M}=\frac{(1+q)^t-1}{q(1+q)^t} \beta C_{INV}
\end{equation}
\begin{equation}
\begin{aligned}
&C_{REL}=C_{ele} \frac{(1+q)^t-1}{q(1+q)^t} H  \biggl( \sum_{r s \in \Psi_C} \xi^{r s}  \sum_{\omega \in \Omega} \delta^\omega \sum_{k \in \Psi_N^{WT}}  P_{k} \zeta^\omega \\
& \frac{\tau_{S W}m_k^{r s}+\tau_{R P}n_k^{r s}}{\tau_{S W}+\tau_{R P}} + \xi^{NO} \sum_{\omega \in \Omega} \delta^\omega \sum_{k \in \Psi_N^{WT}} P_{k} \zeta^\omega\\
& \frac{\tau_{S W}m_k^{NO}+\tau_{R P}n_k^{NO}}{\tau_{S W}+\tau_{R P}} \biggr) 
\end{aligned}
\end{equation}
The objective function consists of three items. The first item represents cable investment expenses $C_{INV}$; the second item denotes operation and maintenance costs $C_{O\&M}$, which is assumed to be several times $C_{INV}$ \cite{chen2016collector}; the last item is reliability cost $C_{REL}$ due to equipment faults. 
The superscript $rs$ represents various system operation scenarios, including the scenario of cable fault ($rs\in\Psi_C$) and normal operation scenario ($rs \in \{N O\}$).
The calculation of reliability cost considers the power curtailment under various system operation scenarios for different wind speeds. Given the mutually exclusive nature of stochastic scenarios, the sum of their probabilities must be equal to 1, i.e., $\sum_{\omega \in \Omega} \delta^\omega=1$, $\sum_{r s \in \Psi_C} \xi^{r s} +\xi^{NO}=1$.
As discussed in the Section II. B, the post-fault network reconfiguration measures are considered. The term $P_{k} \zeta^\omega \frac{\tau_{S W}m_k^{r s}+\tau_{R P}n_k^{r s}}{\tau_{S W}+\tau_{R P}}$ represents the $k$th WT's power curtailment in wind speed scenario $\omega$ considering reconfiguration after the fault of cable $rs$. \par
The constraints are divided into four parts, as shown below. 
\subsubsection{DC power flow and security constraints}
The first part of constraints comes from power flow and device capacity constraints. 

Constraint (5) denotes the power balance. And Constraints (6)-(7) represent DC power flow constraints\cite{shen2023optimal}. The Big-M method is employed in (6), where $M$ is a sufficiently large number. The variable $s_{ij}^{rs}$ indicates whether cable $ij$ is connected after post-fault network reconfiguration following the fault of cable $rs$ (or during normal operation $rs \in \{NO\}$). When cable $ij$ is connected ($s_{ij}^{rs}=1$), Constraint (6) describes the relationship between the phase angle differences $\theta_j^{rs}-\theta_i^{rs}$ and the power flow $P_{i j}^{rs}$ on cable $ij$. When cable $ij$ is disconnected ($s_{ij}^{rs}=0$), Constraint (6) becomes inactive. Constraint (7) specifies that the OSS serves as the slack bus, with its phase angle set to zero. The bi-directional power flow through cable $ij$ is restricted by its connection status, as shown in (8). Constraints (9)-(11) restrict the power flow through the cables and the feeders to not exceed their respective rated capacities. Note that the power losses can be approximately modeled by $P^{rs}_{loss}\approx B_{ij}\left(\theta_j^{rs}-\theta_i^{rs}\right)^2$, as in \cite{yang2017lmp}. However, since this study focuses on reliability enhancement, we only give a brief introduction and do not make more extensive analyses.
\begin{gather}\small
P_{ij}^{rs}=\sum_{k \in \Psi_i} P_{ki}^{rs}+P_i^{rs}, \quad \forall i \in \Psi_N^{WT} \\
\left|B_{i j}\left(\theta_j^{rs}-\theta_i^{rs}\right)-P_{i j}^{rs}\right| \leq\left(1-s_{ij}^{rs}\right) M ,\quad \forall ij \in \Psi_C \\
\theta_j^{rs}=0,\quad \forall j \in \Psi_N\setminus\Psi_N^{WT} \\
-M s_{i j}^{rs} \leq P_{i j}^{rs} \leq M s_{i j}^{rs}, \quad \forall i j \in \Psi_C \\
-P_{i j}^{C} \leq P_{i j}^{rs} \leq P_{i j}^{C}, \quad \forall i j \in \Psi_C \\
P_{f}^{rs}=P_{br^f}^{rs}, \quad \forall f \in \Psi^{F}, \quad br^{f} \in \Psi_C \\
P_{f}^{rs} \leq P_{f}^{C}, \quad \forall f \in \Psi_{F} \\
\forall rs \in \Psi_C \cup\{N O\} \quad \text{for \quad (5)-(11)} \nonumber
\end{gather}
\subsubsection{Fault impact identification constraints} 
The second set of constraints is used to identify the minimum outage area caused by faults, which depends on the normal operation state. 
When one cable fails, the CB on the feeder to which the faulty cable belongs will trip, preventing all the WTs connected to that feeder from supplying power to the OSS.\par

Constraint (12) serves to identify the outage propagation area.
$h_{k}^{f}$ ($h_{ij}^{f}$) represents the affiliation relationship between WT $k$ (cable $ij$) and the feeder $f$ during normal operation, while $m_{k}^{rs}$ indicates whether WT $k$ is affected by the fault of cable $rs$. 
If WT $k$ and cable $rs$ are connected to the same OSS through the same feeder $f$ during normal operation ($h_{k}^{f}=h_{rs}^{f}=1$), WT $k$ will experience a power outage due to the fault in cable $rs$ ($m_{k}^{rs}=1$).
The affiliation relationship is propagated through connected cables, as shown in (13)-(14). When cable $ij$ is connected during normal operation ($s_{ij}^{NO}=1$), WTs $i$ and $j$, as well as cable $ij$, are considered affiliated to the same feeder ($h_{i}^{f}=h_{j}^{f}=h_{ij}^{f}$). Constraints (15)-(16) define the source of the affiliation relationship, and (17)-(18) ensure each WT or cable is affiliated to at most one feeder.
\vspace{-0.5em}
\begin{gather}
h_{k}^{f}+h_{rs}^{f}-1 \leq m_{k}^{rs}, \forall f \in \Psi_F, \forall k \in \Psi_N^{WT} \\
\begin{aligned}
\left| h_{i j}^{f}-h_{i}^{f}\right| \leq M\left(1-s_{i j}^{NO}\right), & \forall ij \in \Psi_C, \forall f \in \Psi_F
\end{aligned}\\
\begin{aligned}
\left| h_{i j}^{f}-h_{j}^{f}\right| \leq M\left(1-s_{i j}^{NO}\right), & \forall ij \in \Psi_C, \forall f \in \Psi_F
\end{aligned} \\
h_{br^f}^{f}=s_{br^f}^{NO},\forall f \in \Psi_F, br^f \in \Psi_C \\
h_{ij}^{f} \leq s_{ij}^{NO},\forall f \in \Psi_F, \forall ij \in \Psi_C \\
\sum_{f \in \Psi_F} h_{k}^{f} \leq 1, \quad \forall k \in \Psi_N^{WT} \\
\sum_{f \in \Psi_F} h_{ij}^{f} \leq 1, \quad \forall i j \in \Psi_C 
\end{gather}
\vspace{-0.5em}
\subsubsection{Post-fault network reconfiguration constraints}
The third set of constraints aims to determine optimal post-fault network reconfiguration strategies to maximize reliability.\par

Equation (19) ensures the unavailability of the faulty cable, and Cons. (20) states that the connectivity of the cable is subject to its installation.
The connection status is set to zero ($s_{ij}^{rs}=0$) when no cable is invested in ($l_{ij}=0$).
Constraint (21) specifies that the WTs not affected by the fault should maintain power supply after the network reconfiguration. Specifically, for any WT $k$, if $m^{rs}_k=0$, then $n^{rs}_k=0$.
Equations (22) and (24) represent WT's output under fault and normal operation conditions, respectively. Equation (23) ensures ECS operates in a radial structure. The wind curtailment of WTs under normal conditions is supposed to be zero, as described in (24)-(25).
\vspace{-0.5em}
\begin{gather}
s_{rs}^{rs}=0 \label{fault impact} \\
s_{ij}^{rs} \leq l_{ij}, \forall ij \in \Psi_{C} \\
m^{rs}_k \geq n^{rs}_k, \forall k \in \Psi_{N}^{WT} \\
P_{k}^{rs}=P_{k}\left(1-n_{k}^{rs}\right), \forall k \in \Psi_{N}^{WT} \\
\sum_{i j \in \Psi_C}s_{ij}^{rs}=\sum_{k \in \Psi_{N}^{W T}}\left(1-n_{k}^{rs}\right) \label{radial} \\
\forall rs \in \Psi_C \quad \text{for \quad (19)-(23)} \nonumber \\
P_{k}^{NO}=P_k, \forall k \in \Psi_{N}^{WT} \\
m_{k}^{NO}=n_{k}^{NO}=0, \forall k \in \Psi_{N}^{WT}
\end{gather}
\subsubsection{Non-crossing constraints}
The fourth part incorporates geographical constraints, aiming to identify crossing cable pairs via (26) and avoid infeasible layouts by imposing (27).
\begin{gather}
\begin{aligned}
&\mathcal{X}_C=\{(i_1 j_1, i_2 j_2) \mid(\overrightarrow{i_1 i_2} \times \overrightarrow{i_1 j_1}) \cdot(\overrightarrow{i_1 j_2} \times \overrightarrow{i_1 j_1})<0 \\
& \land(\overrightarrow{i_2 j_1} \times \overrightarrow{i_2 j_2}) \cdot(\overrightarrow{i_2 i_1} \times \overrightarrow{i_2 j_2})<0, \forall i_1j_1, i_2j_2 \in \Psi_C \}
\end{aligned}\\
l_{i_1j_1}+l_{i_2j_2} \leq 1,\quad \forall \left(i_1j_1, i_2j_2 \right) \in \mathcal{X}_C 
\end{gather}
Therefore, the proposed model is formulated as:
$$
\begin{gathered}
\quad \min \text{(1)} \\
s.t. \text { (2)-(27)}.
\end{gathered}
$$



\section{Solution Framework}
In the reliability-based ECS cable layout planning problem, the number of candidate cables far exceeds the actual installations in the final cable layout. That means many contingency scenarios for non-installed cables only increase the problem's complexity.
Meanwhile, we notice that the renowned Column-and-Constraint Generation (CCG) algorithm \cite{zeng2013solving}, designed for solving two-stage robust optimization problems, leverages a formulation based on partial enumeration within a subset of the scenario set. This approach provides a valid relaxation. The CCG procedure strategically identifies and gradually incorporates non-trivial scenarios into the subset. This gradual incorporation serves to tighten the relaxation until convergence.
Inspired by the insights, we propose an iterative solution framework. We begin by establishing the theoretical foundation and then provide a detailed description of the algorithm's workflow. 
\vspace{-1em}
\subsection{Theoretical Foundation}
The reliability-based cable layout planning problem proposed in Section III is defined as $\mathcal{P}_{\Upsilon,\Omega}(x,y^{\upsilon,\omega})$. Its compact form can be expressed as follows.
\begin{subequations}
\begin{align}
& \begin{aligned}
&\mathcal{P}_{\Upsilon,\Omega}(x,y^{\upsilon,\omega}) = \min _{x, y^{\upsilon,\omega}} c^Tx+\sum_{\substack{\upsilon \in \Upsilon \\ \upsilon\neq \upsilon_0}} \xi^\upsilon \sum_{\omega \in \Omega}\delta^\omega  d^Ty^{\upsilon,\omega} \\ 
&+(1-\sum_{\substack{\upsilon \in \Upsilon \\ \upsilon\neq \upsilon_0}} \xi^\upsilon ) \sum_{\omega \in \Omega} \delta^\omega  d^Ty^{\upsilon_0,\omega}
\end{aligned} \\
&s.t. \nonumber \\
&\underline{b}\leq Ax\leq \bar{b} \\
&\underline{h}^{\upsilon,\omega}\leq Tx+W^{\upsilon,\omega}y^{\upsilon,\omega}\leq \bar{h}^{\upsilon,\omega} \\
&\underline{e}^{\upsilon,\omega}\leq D^{\upsilon,\omega}y^{\upsilon,\omega}\leq \bar{e}^{\upsilon,\omega} \\
&x \in \{0,1\}^n, y^{\upsilon,\omega} \geq 0 \\
&\forall \upsilon \in \Upsilon, \forall \omega \in \Omega \text{ for (28c)-(28e)} \nonumber
\end{align}
\end{subequations}
Here, $x$ represents the first-stage variables and corresponds to $\{l_{ij}\}$ in the proposed model.
$y^{\upsilon,\omega}$ denotes the second-stage variables associated with ECS operational state scenario $\upsilon$ and wind speed scenario $\omega$. The second-stage variables encompass $\{s_{ij}^{NO/rs}, h_{k}^f, h_{ij}^f, m_{k}^{NO/rs}, n_{k}^{NO/rs}, P_{f}^{rs}, P_{ij}^{rs}, P_{k}^{rs}, \theta_{i}^{rs}\}$. Clearly, the inclusion of binary variables makes the problem less straightforward to solve.
$\Upsilon$ is the operational state scenario set, and $\upsilon_0$ denotes the normal operation scenario of ECS. $\Omega$ defines the wind speed scenario set.
$\xi^\upsilon$ represents the probability of ECS operation scenario $\upsilon$, while $\delta^\omega$ denotes the probability of wind speed scenario $\omega$.
$c$ and $d$ are the cost coefficients related to first-stage and second-stage decisions in the objective function (28a).
$A$, $T$, $W^{\upsilon,\omega}$, and $D^{\upsilon,\omega}$ are coefficient matrices used to enforce the first-stage constraint (28b), coupling constraints (28c), and second-stage constraint (28d).
$\underline{b}$ and $\bar{b}$ define bounds on $Ax$, while $\underline{h}^{\upsilon,\omega}$ and $\bar{h}^{\upsilon,\omega}$, as well as $\underline{e}^{\upsilon,\omega}$ and $\bar{e}^{\upsilon,\omega}$, define bounds on $Tx+W^{\upsilon,\omega}y^{\upsilon,\omega}$ and $D^{\upsilon,\omega}y^{\upsilon,\omega}$, respectively.\\
\textbf{Definition 1.} 
A problem with \textit{contingency structure} \cite{lumbreras2012progressive} is defined as a stochastic programming problem where uncertainty arises from the faults of installed components. And the operational cost of any contingency scenario should be higher than or equal to the cost of the normal operation scenario. \par 
Clearly, the proposed reliability-based ECS cable layout planning problem naturally fits \textbf{Definition 1}. \\
\textbf{Remark 1.} 
According to Assumption 2), the second-stage constraints describe the faulty operational states for all candidate cables as well as the normal operation state. Given the vast number of candidate cables, the inclusion of operational state scenarios inevitably leads to a significant increase in the problem's complexity.\\
\textbf{Remark 2.} If one candidate cable is not invested for installation in the first stage, its corresponding faulty operational state scenario is deemed irrelevant. In other words, the associated objective function terms and second-stage constraints should not influence the problem resolution. Therefore, the faulty operational scenarios of non-installed cables have the same impact as the normal operation scenario.
\par
By fixing the value of $x$ as $\hat{x}$ in the feasible set, the \textit{recourse problem} $\mathcal{Q}_{\Upsilon,\Omega}(\hat{x})$ can be obtained as follows. 
\begin{subequations}
\begin{align}
&\begin{aligned}
&\mathcal{Q}_{\Upsilon,\Omega}(\hat{x}) \\
& =\min _{y^{\upsilon,\omega}}\sum_{\substack{\upsilon \in \Upsilon \\ \upsilon\neq \upsilon_0}} \xi^\upsilon \sum_{\omega \in \Omega}\delta^\omega  d^Ty^{\upsilon,\omega} 
+\small(1-\sum_{\substack{\upsilon \in \Upsilon \\ \upsilon\neq \upsilon_0}} \xi^\upsilon\small) \sum_{\omega \in \Omega} \delta^\omega  d^Ty^{\upsilon_0,\omega} \\
& =\min _{y^{\upsilon,\omega}}\sum_{\upsilon \in \Upsilon} \xi^\upsilon \sum_{\omega \in \Omega}\delta^\omega  d^Ty^{\upsilon,\omega}
\end{aligned} \\
&s.t. \nonumber \\
&\underline{h}^{\upsilon,\omega}-T\hat{x}\leq W^{\upsilon,\omega}y^{\upsilon,\omega}\leq \bar{h}^{\upsilon,\omega}-T\hat{x} \\
&\underline{e}^{\upsilon,\omega}\leq D^{\upsilon,\omega}y^{\upsilon,\omega}\leq \bar{e}^{\upsilon,\omega} \\
&y^{\upsilon,\omega} \geq 0 \\
& \forall \upsilon \in \Upsilon, \forall \omega \in \Omega \text{ for (29b)-(29d)} \nonumber
\end{align}
\end{subequations}
\par
For better illustration, let $y^{\upsilon,\omega *}(\hat{x})$ represent the optimal solution of $\mathcal{Q}_{\Upsilon,\Omega}(\hat{x})$ for operational state scenario $\upsilon$ and wind speed scenario $\omega$:
\begin{equation}
\begin{aligned}
y^{\upsilon,\omega *}(\hat{x})=\underset{y^{\upsilon,\omega}(\hat{x}) \in \mathcal{Y}^{\upsilon,\omega}(\hat{x})}{\arg\min}\sum_{\upsilon \in \Upsilon} \xi^\upsilon \sum_{\omega \in \Omega}\delta^\omega  d^Ty^{\upsilon,\omega}
\end{aligned}
\end{equation}
where $\mathcal{Y}^{\upsilon,\omega}(\hat{x})$ denotes the feasible region restricted by constraints (29b)-(29d). According to \textbf{Definition 1}, it can be derived that, 
\begin{equation}
d^T y^{\upsilon,\omega *}(\hat{x}) \geq d^T y^{\upsilon_0,\omega *}(\hat{x}), \forall \upsilon \in \Upsilon, \forall \omega \in \Omega.
\end{equation}
According to \textbf{Remark 1}, the computational burden of solving the \textit{primal problem} $\mathcal{P}_{\Upsilon,\Omega}(x,y^{\upsilon,\omega})$ can be substantial and may even lead to unsolvability. Hence, it is necessary to make some simplifications to the problem.
It can be referred from \textbf{Remark 2} that, if the $i$th candidate cable is not invested for installation ($\hat{x}_i=0$), then the optimal solution $y^{\upsilon_i,\omega *}$ in the faulty operational scenario $\upsilon_i$ is equal to the optimal solution $y^{\upsilon_0,\omega *}$ in the normal operation scenario $\upsilon_0$. This is given by,
\begin{equation}
\left|y^{\upsilon_i,\omega *}(\hat{x})-y^{\upsilon_0,\omega *}(\hat{x})\right| \leq M\hat{x}_i, \forall \omega \in \Omega.
\end{equation}
(32) shows that, only the faulty operational scenarios related to those installed cables matter. And we can equivalently treat the contingency scenarios of non-installed cables as the normal operation scenario. This is used to simplify the problem.\par
Firstly, a \textit{simplified problem} $\mathcal{\Tilde{P}}_{\Tilde{\Upsilon},\Omega}(x,y^{\upsilon,\omega})$ of the \textit{primal problem} $\mathcal{P}_{\Upsilon,\Omega}(x,y^{\upsilon,\omega})$ is defined by only considering a subset $\Tilde{\Upsilon}$ of the system operational state scenario set $\Upsilon$ in the second stage. As for the other scenarios, we temporarily consider them as normal operational state scenarios.
\begin{subequations}
\begin{align}
&\begin{aligned}
&\mathcal{\Tilde{P}}_{\Tilde{\Upsilon},\Omega}(x,y^{\upsilon,\omega}) = \min _{x, y^{\upsilon,\omega}} c^Tx+\sum_{\substack{\upsilon \in \Tilde{\Upsilon} \\ \upsilon\neq \upsilon_0}} \xi^\upsilon \sum_{\omega \in \Omega}\delta^\omega  d^Ty^{\upsilon,\omega} \\ 
&+(1-\sum_{\substack{\upsilon \in \Tilde{\Upsilon} \\ \upsilon\neq \upsilon_0}} \xi^\upsilon ) \sum_{\omega \in \Omega} \delta^\omega  d^Ty^{\upsilon_0,\omega}
\end{aligned}\\
&s.t.\quad  \text{(28b)-(28e)}, \forall \upsilon \in \Tilde{\Upsilon}, \forall \omega \in \Omega \text{ for (28c)-(28e)}
\end{align}
\end{subequations}
And the corresponding \textit{simplified recourse problem} $\mathcal{\Tilde{Q}}_{\Tilde{\Upsilon},\Omega}(\hat{x})$ for a given feasible solution $\hat{x}$ is given by (34).
\begin{subequations}
\begin{align}
&\begin{aligned}
&\mathcal{\Tilde{Q}}_{\Tilde{\Upsilon},\Omega}(\hat{x})\\
& =\min _{y^{\upsilon,\omega}}\sum_{\substack{\upsilon \in \Tilde{\Upsilon} \\ \upsilon\neq \upsilon_0}} \xi^\upsilon \sum_{\omega \in \Omega}\delta^\omega  d^Ty^{\upsilon,\omega} 
+\small(1-\sum_{\substack{\upsilon \in \Tilde{\Upsilon} \\ \upsilon\neq \upsilon_0}} \xi^\upsilon\small) \sum_{\omega \in \Omega} \delta^\omega  d^Ty^{\upsilon_0,\omega}
\end{aligned} \\
&s. t. \quad  \text{(29b)-(29d)}, \forall \upsilon \in \Tilde{\Upsilon}, \forall \omega \in \Omega \text{ for (29b)-(29d)}
\end{align}
\end{subequations}
\textbf{Remark 3.} 
As we only consider a subset of the system operational state scenario set($\small\Tilde{\Upsilon}\subseteq\Upsilon\small$), the second-stage constraints in $\mathcal{\Tilde{P}}_{\Tilde{\Upsilon},\Omega}(x,y^{\upsilon,\omega})$ are greatly reduced, which makes it easier to solve.
Furthermore, since we treat some faulty state scenarios as normal state scenarios, combined with (31), it follows that the optimal value of  $\mathcal{\Tilde{Q}}_{\Tilde{\Upsilon},\Omega}(\hat{x})$ is necessarily lower than or equal to the optimal value of $\mathcal{Q}_{\Upsilon,\Omega}(\hat{x})$. Consequently, the optimal value of $\mathcal{\Tilde{P}}_{\Tilde{\Upsilon},\Omega}(x,y^{\upsilon,\omega})$ is not larger than $\mathcal{P}_{\Upsilon,\Omega}(x,y^{\upsilon,\omega})$.\\
\textbf{Remark 4.} 
For any given feasible solution $\hat{x}$, define $\hat{\Upsilon}$ as the set that includes the faulty operational state scenarios corresponding to all the installed cables and the normal state scenario, $\hat{\Upsilon}=\{\upsilon_i|\upsilon_i \in \Upsilon, \hat{x}_i=1\} \cup \upsilon_0$.
If all the scenarios that describe the faults of installed cables are included in the reduced set, i.e., $\hat{\Upsilon} \subseteq \Tilde{\Upsilon}$, then the optimal value of  $\mathcal{\Tilde{Q}}_{\Tilde{\Upsilon},\Omega}(\hat{x})$ must be equal to $\mathcal{Q}_{\Upsilon,\Omega}(\hat{x})$. The proof is given below.\par


In the \textit{simplified recourse problem} $\mathcal{\Tilde{Q}}_{\Tilde{\Upsilon},\Omega}(\hat{x})$, if $\hat{\Upsilon} \subseteq \Tilde{\Upsilon}$, then set $\Tilde{\Upsilon}$ can be divided into $\hat{\Upsilon}$ and the remaining part $\Tilde{\Upsilon}\setminus\hat{\Upsilon}$. According to (32), the scenarios of non-installed cables ($\upsilon \in \Tilde{\Upsilon}\setminus\hat{\Upsilon}$) have the same impact as normal state scenario $\upsilon_0$. 
So $\mathcal{\Tilde{Q}}_{\Tilde{\Upsilon},\Omega}(\hat{x})$ in (34a) can be rewritten as follows.
\begin{equation}
\begin{aligned}
&\mathcal{\Tilde{Q}}_{\Tilde{\Upsilon},\Omega}(\hat{x})\\
& =\min _{y^{\upsilon,\omega}}\sum_{\substack{\upsilon \in \hat{\Upsilon}  \\ \upsilon\neq \upsilon_0}} \xi^\upsilon \sum_{\omega \in \Omega}\delta^\omega  d^Ty^{\upsilon,\omega}
+\sum_{\substack{\upsilon \notin \hat{\Upsilon} \\ 
\upsilon \in \Tilde{\Upsilon}}} \xi^\upsilon \sum_{\omega \in \Omega}\delta^\omega  d^Ty^{\upsilon_0,\omega} \\
& +\small(1-\sum_{\substack{\upsilon \in \Tilde{\Upsilon} \\ \upsilon\neq \upsilon_0}} \xi^\upsilon\small) \sum_{\omega \in \Omega} \delta^\omega  d^Ty^{\upsilon_0,\omega} \\
& =\min _{y^{\upsilon,\omega}}\sum_{\substack{\upsilon \in \hat{\Upsilon} \\ \upsilon\neq \upsilon_0}} \xi^\upsilon \sum_{\omega \in \Omega}\delta^\omega  d^Ty^{\upsilon,\omega}
+\small(1-\sum_{\substack{\upsilon \in \hat{\Upsilon} \\ \upsilon\neq \upsilon_0}} \xi^\upsilon\small) \sum_{\omega \in \Omega} \delta^\omega  d^Ty^{\upsilon_0,\omega}
\end{aligned}
\end{equation}
It can be seen that the optimal value of the simplified recourse problem $\mathcal{\Tilde{Q}}_{\Tilde{\Upsilon},\Omega}(\hat{x})$ is solely determined by $\hat{\Upsilon}$. Hence, for all the recourse problems satisfying the condition $\hat{\Upsilon} \subseteq \Tilde{\Upsilon}$, their optimal values are the same and equal to the optimal value of the complete recourse problem $\mathcal{Q}_{\Upsilon,\Omega}(\hat{x})$.
Let $(x^*,y^{\upsilon,\omega *})$ be the optimal solutions of the \textit{simplified problem} $\mathcal{\Tilde{P}}_{\Tilde{\Upsilon},\Omega}(x,y^{\upsilon,\omega})$, 
\begin{equation}
(x^*,y^{\upsilon,\omega *}) \equiv \underset{x,y^{\upsilon,\omega}}{\arg\min} \mathcal{\Tilde{P}}_{\Tilde{\Upsilon},\Omega}(x,y^{\upsilon,\omega}), \upsilon \in \Tilde{\Upsilon}, \omega \in \Omega.
\end{equation}
\textbf{Definition 2.} The \textit{complete incorporation condition} is defined as follows: all faulty state scenarios related to the installed cables in the optimal solution $x^*$ are included in the reduced operational state scenario set of the simplified problem, which is given by $\Upsilon^* \subseteq \Tilde{\Upsilon}, \text{where } \Upsilon^*=\{\upsilon_i|\upsilon_i \in \Upsilon, x^*_i=1\} \cup \upsilon_0$.\\
\textbf{Remark 5.} Once the \textit{complete incorporation condition} is met, the optimal solution to the \textit{simplified problem} $\mathcal{\Tilde{P}}_{\Tilde{\Upsilon},\Omega}(x,y^{\upsilon,\omega})$ precisely aligns with the optimal solution to the \textit{primal problem} $\mathcal{P}_{{\Upsilon},\Omega}(x,y^{\upsilon,\omega})$. The proof can be found in the Appendix. \par
\vspace{-1em}

\subsection{Customized Progressive Contingency Incorporation}
Motivated by \textbf{Remark 5}, we propose an iterative framework to seek and solve the \textit{simplified problem} that satisfies the \textit{complete incorporation condition}, leading to the optimal solution of the \textit{primal problem}. Here, we introduce the customized progressive contingency incorporation (PCI) (Algorithm \ref{alg:CPCI}), integrating the Benders Decomposition (BD) strategy \cite{benders2005partitioning}. \par
\floatname{algorithm}{Algorithm}
\renewcommand{\algorithmicrequire}{\textbf{Initialization:}} 
\renewcommand{\algorithmicensure}{\textbf{Iteration:}}
\begin{algorithm}[htbp]\small   
\caption{ Customized PCI (CPCI) algorithm }    
\label{alg:CPCI} 
\begin{algorithmic}[1]                
\Require  \\ 
$\Tilde{\Upsilon}=\upsilon_0, \Omega=\omega_n$;\\
Apply BD strategy to solve  $(\hat{x},\hat{y}^{\upsilon,\omega}) =\underset{x,y^{\upsilon,\omega}}{\arg\min} \mathcal{\Tilde{P}}_{\Tilde{\Upsilon},\Omega}$ to $\epsilon \leq \epsilon_0$;\\
$\hat{\Upsilon}=\{\upsilon_i|\upsilon_i \in \Upsilon \cup \upsilon_0, \hat{x}_i=1\}$, $x_{ws}=\hat{x}$, $Ind=0$; 
\Ensure  
\While    {$Ind==0$}
\State $\Tilde{\Upsilon}=\Tilde{\Upsilon} \cup \hat{\Upsilon}$;
\State Apply BD strategy to solve $(\hat{x},\hat{y}^{\upsilon,\omega}) =\underset{x,y^{\upsilon,\omega}}{\arg\min} \mathcal{\Tilde{P}}_{\Tilde{\Upsilon},\Omega}$ to $\epsilon \leq \epsilon_0$ with warm-start point $x_{ws}$;
\State $\hat{\Upsilon}=\{\upsilon_i|\upsilon_i \in \Upsilon, \hat{x}_i=1\}\cup \upsilon_0$, $x_{ws}=\hat{x}$;
\If{$\hat{\Upsilon}==\hat{\Upsilon} \cap \Tilde{\Upsilon}$}
\State Apply BD strategy to solve $(x^*,y^{\upsilon,\omega *})=\underset{x,y^{\upsilon,\omega}}{\arg\min}\mathcal{\Tilde{P}}_{\Tilde{\Upsilon},\Omega}$ to \textit{optimality} with warm-start point $x_{ws}$;
\State $\Upsilon^*=\{\upsilon_i|\upsilon_i \in \Upsilon, x_i^*=1\}\cup \upsilon_0$;
\If{$\Upsilon^*==\Upsilon^* \cap \Tilde{\Upsilon}$}
\State $Ind=1$;
\Else
\State $\Tilde{\Upsilon}=\Tilde{\Upsilon} \cup \Upsilon^*$, $x_{ws}=x^*$;
\EndIf
\EndIf
\EndWhile
\end{algorithmic}
\end{algorithm}
\setlength{\intextsep}{1pt} 

\begin{figure*}[!htbp]
\centering
\includegraphics[width=\textwidth]{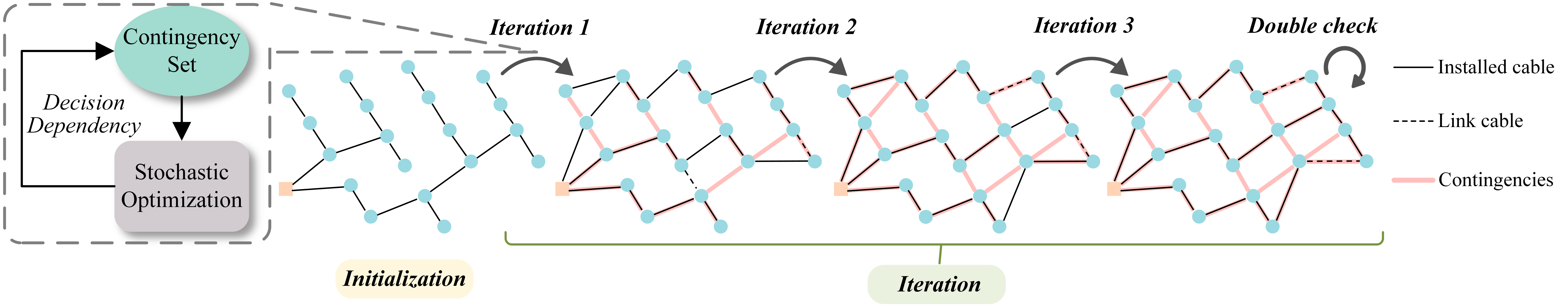}
\vspace{-2em}
\caption{The illustrative diagram of the CPCI algorithm. }
\label{CPCI}
\vspace{-1.5em}
\end{figure*}

First, the algorithm initializes $\Tilde{\Upsilon}$ with $\upsilon_0$ and $\Omega$ with nominal wind scenario $\omega_n$.
Employing the BD strategy, the algorithm tackles the \textit{simplified problem} $\mathcal{\Tilde{P}}_{\Tilde{\Upsilon},\Omega}(x,y^{\upsilon,\omega})$ to derive a decent solution $\hat{x}$ with a termination criterion of $\epsilon \leq \epsilon_0$. Here, $\epsilon_0$ denotes the CPCI optimality gap tolerance. 
More specifically, the standard BD strategy is utilized to further decompose the \textit{simplified problem}. The problem is partitioned into two components: (i) a master problem that includes all integer variables, thus forming an integer programming problem, and (ii) a subproblem that contains all continuous variables, formulated as a linear programming problem. The master problem and subproblem are solved iteratively. The master problem determines the integer variables and passes them into the subproblem. Subsequently, ``Benders optimality cuts'' or ``Benders feasibility cuts'' generated from the subproblem are fed back into the master problem to refine the solution for subsequent iterations. Further details on BD strategy can be found in \cite{benders2005partitioning,taskin2010benders}.
$\hat{\Upsilon}$ and the warm-start point $x_{ws}$ are set based on $\hat{x}$, and the indicator for the \textit{complete incorporation condition} is initialized to 0 (lines 1-3). 
Next, the algorithm iterates to solve $\mathcal{\Tilde{P}}_{\Tilde{\Upsilon},\Omega}$ and gradually incorporates the related operational state scenarios $\hat{\Upsilon}$ into $\Tilde{\Upsilon}$ (lines 5-7). The algorithm then checks if the \textit{complete incorporation condition} is satisfied (lines 8-16). If the condition holds, as stated in \textbf{Remark 5}, the obtained solution $(x^*, y^{\upsilon,\omega *})$ stands as the global optimal solution, and the iteration is terminated. Otherwise, the algorithm continues to make another attempt in the next iteration. Fig. \ref{CPCI} shows the diagram of CPCI.

The Traditional PCI (TPCI) algorithm \cite{lumbreras2012progressive} exhibits efficient performance in early iterations. However, as the iterations progress, the addition of scenarios leads to increased time for solving the simplified problem to optimality. After several trials, it is observed that the optimal solutions are often identified early in the optimization process, with a substantial amount of time dedicated to verifying their optimality. To this end, we introduce a termination criterion during the optimization process of each iteration. 
To obtain the globally optimal solution, a double-check mechanism (lines 11-15) is employed for final verification. If the reduced scenario set $\Tilde{\Upsilon}$ cannot remain stable during the iterations, it will expand until all contingencies are incorporated. In this extreme case, the algorithm still converges and achieves the optimal solution. CPCI algorithm is devised to strike a better balance between efficiency and accuracy in the pursuit of optimal solutions.
\vspace{-.5em}
\section{Case Study}
\vspace{-.2em}
In this section, we start by validating the proposed model on a 30-WT OWF used in \cite{shen2023optimal}. Then, we demonstrate the adaptability of our method by solving Race Bank OWF with 91 WTs. Finally, we apply different solution frameworks to Triton OWF to compare their efficiency. 
Simulations were conducted on a laptop PC equipped with an Intel Core i5 processor using CPLEX 12.10.0. The information of WTs and OSSs are obtained from \cite{kisorca}. 
To make this paper concise, the complete datasets for the case studies are available in \cite{Ding2023}.
\vspace{-1em}
\subsection{Effectiveness Validation of the Proposed Method}
We utilize the 30-WT OWF in \cite{shen2023optimal} as the first benchmark. 
Cases 1-4 are set up to test the effectiveness of the proposed planning method.\\
\textbf{Case 1:} ECS planning with radial structural limitation \cite{shen2021large};\\
\textbf{Case 2:} ECS planning without predefined structural limitation (the proposed method);\\
\textbf{Case 3:} ECS planning with multi-loop structural limitation\cite{gong2017optimal};\\
\textbf{Case 4:} ECS planning with ring structural limitation\cite{shen2023optimal}.\par
\begin{figure}[!htbp]
\centering
\includegraphics[width=0.5\textwidth]{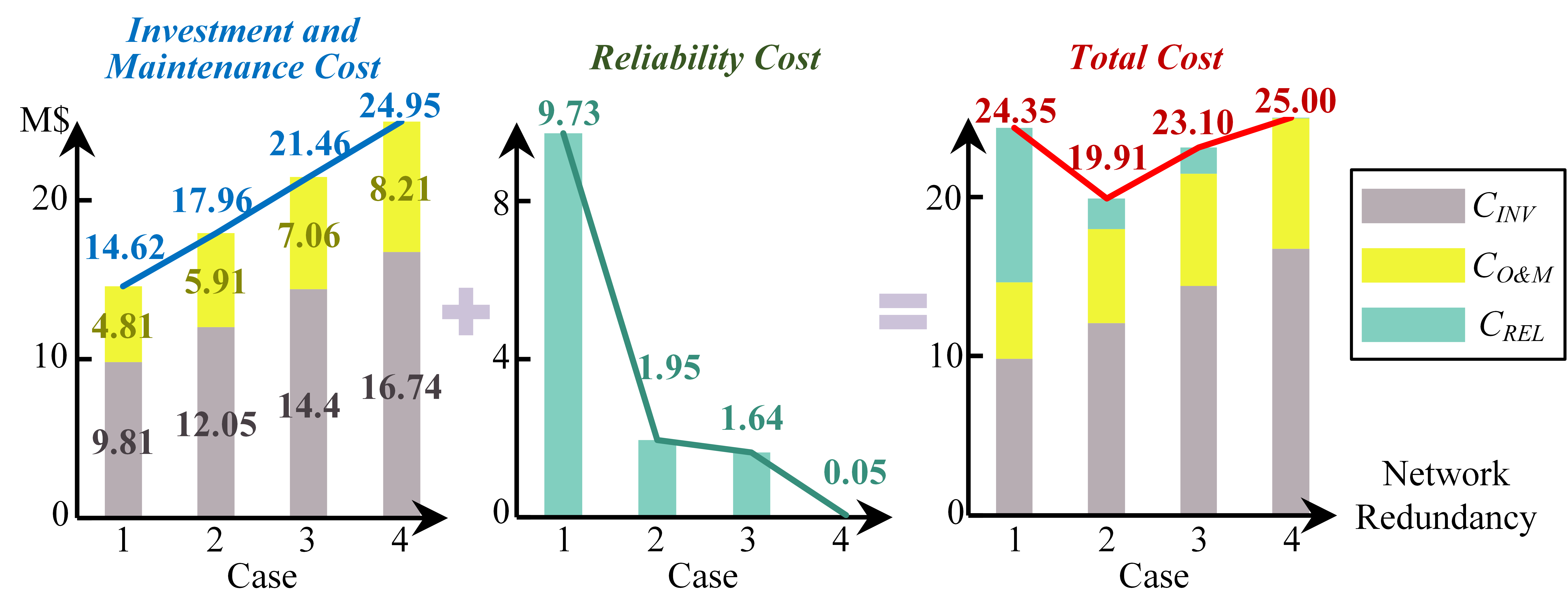}
\vspace{-2em}
\caption{30-WT ECS cable layout planning results.}
\label{Cost-Rel}
\end{figure}
\begin{figure*}[!htbp]
\centering
\includegraphics[width=\textwidth]{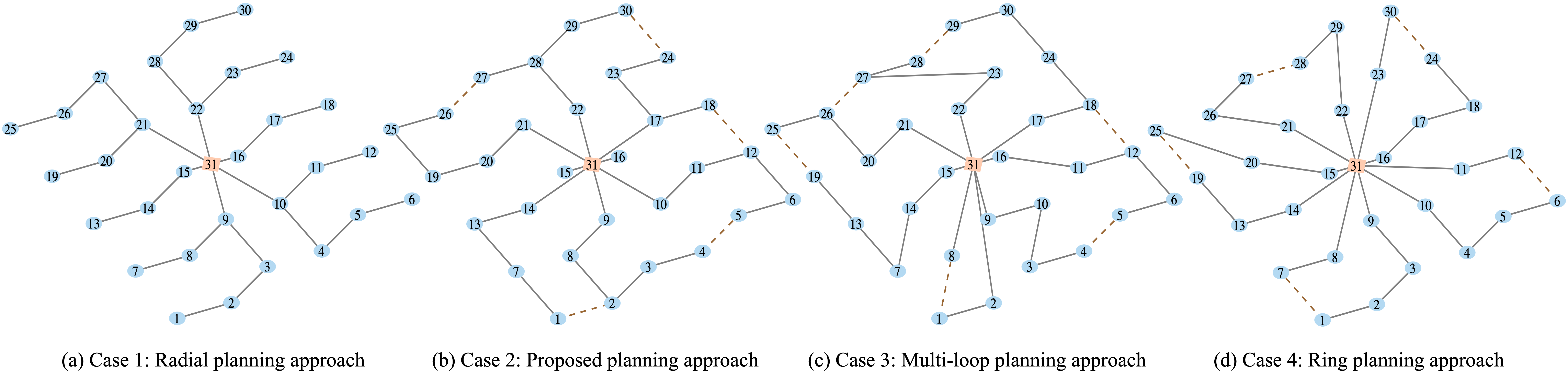}
\vspace{-2em}
\caption{Cable layouts in (a) Case 1: Radial planning approach (b) Case 2: Proposed planning approach (c) Case 3: Multi-loop planning approach (d) Case 4: Ring planning approach for 30-WT ECS. }
\label{30WT}
\vspace{-1em}
\end{figure*}

The planning results and cable layout solutions are shown in Fig. \ref{Cost-Rel} and Fig. \ref{30WT}. The correlation between total cost and network redundancy aligns with the cost-benefit balance curve in \cite{djapic2008cost}, as depicted by the red curve in Fig. \ref{Cost-Rel}. 
In this study, the wind conditions are assumed to remain stable annually, and typical wind scenarios are utilized to represent the wind conditions throughout the lifespan of the OWF. Thus, it is important to note that the results shown are in-sample results.\par
The investment and maintenance costs for ECS significantly increase with the extension of cable length. The total cable lengths of Cases 1-4 are 32.60 km, 40.03 km, 47.86 km, and 55.63 km, respectively.
Obviously, the radial structure in Fig. \ref{30WT}(a) excels in this regard. 
However, the long-term wind curtailment cost, i.e., the reliability cost, is another pivotal factor in assessing ECS design. 
The findings indicate that the radial structure in Case 1, without any network redundancy, provides economic advantages but incurs substantial energy loss during faults. Its reliability cost accounts for up to 40\% of the total costs.
Conversely, the double-sided ring structure in Case 4 presents significant potential for improving ECS reliability. Nevertheless, an overemphasis on reliability enhancement leads to a substantial rise in cable-related costs. The benefits gained from reliability enhancement are offset by the increased investment required in this OWF. 
Therefore, the optimal ECS must balance economic efficiency and reliability effectively. Clearly, the conventional planning approach used in Cases 1, 3 and 4, constrained by predefined topological restrictions, is insufficient to realize this objective.\par
The proposed method used in Case 2 demonstrates its capability to identify the optimal ECS layout, as depicted in Fig. \ref{30WT}(b).
In Case 2, the overall cost of the optimal structure reduces by around 18.2\%, 13.8\%, and 20.4\% compared to Cases 1, 3, 4. The installation of link cables results in a 22.8\% increase in cable length compared to the radial structure. However, there is a significant 80.0\% reduction in reliability cost, attributed to the improved capability for post-fault network reconfiguration. Furthermore, as illustrated in Fig. \ref{Cost-Rel}, the marginal benefits of reliability improvements, brought about by network redundancy, diminish with further increases in cable length.
\par
It should be noted that the optimal structure in Fig. \ref{30WT}(b) shares some similarities with the multi-loop structure in Fig. \ref{30WT}(c), while also incorporating radial structure to optimize ECS's overall cost. However, Case 3 employs a two-step heuristic algorithm \cite{gong2017optimal}, so the optimality of solutions is not guaranteed. Contrarily, the proposed method directly obtains optimal layouts and is applicable to large-scale ECSs, as illustrated in the upcoming subsections.

\subsection{Scalability Validation of the Proposed Method}
To verify the scalability of the proposed method and contrast it with prevalent heuristic methods, we conducted more tests on Race Bank OWF, which consists of 2 OSSs and 91 WTs. Four cases are designed as follows: \\
\textbf{Case 5:} ECS planning with two-phase CWS algorithm \cite{zuo2021two};\\
\textbf{Case 6:} ECS planning with CSI structure \cite{zuo2019collector};\\
\textbf{Case 7:} Proposed ECS planning without OSS interconnection;\\
\textbf{Case 8:} Proposed ECS planning with OSS interconnection.\par
\begin{figure*}[htbp]
\centering
\includegraphics[width=0.78\textwidth]{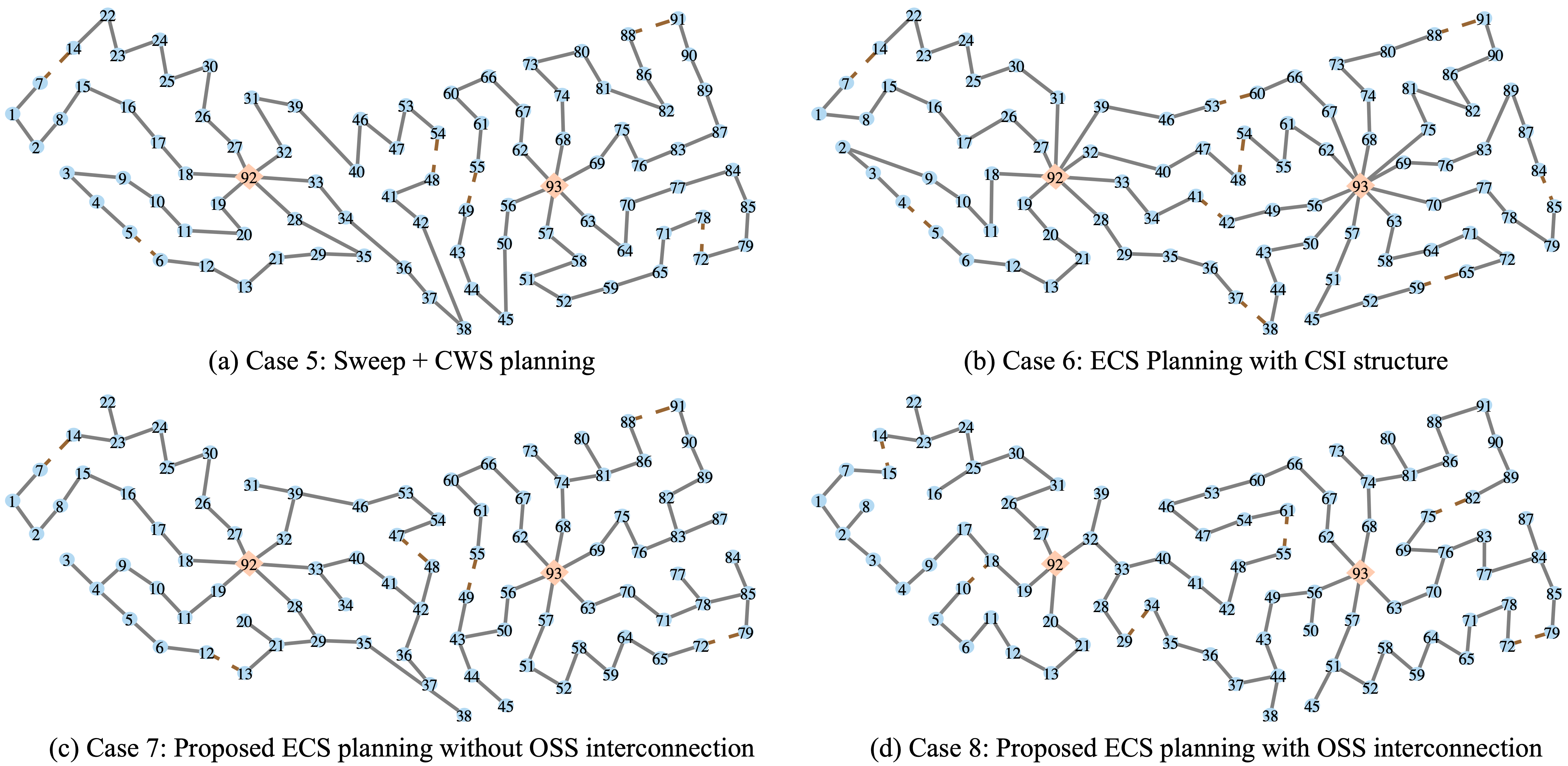}
\vspace{-1em}
\caption{Cable layouts in (a) Case 5: Two-phase CWS algorithm \cite{zuo2021two} (b) Case 6: ECS planning with CSI structure \cite{zuo2019collector} (c) Case 7: Proposed ECS planning without OSS interconnection (c) Case 8: Proposed ECS planning with OSS interconnection for Race Bank OWF. }
\label{91WT}
\vspace{-1.5em}
\end{figure*}

The two-phase algorithm\cite{zuo2021two} in Case 5 involves WT clustering by the sweeping process and cable routing by heuristic CWS algorithm, which has been proven to be effective in many works \cite{zuo2020collector,zuo2021two}.
Case 6 also applies the modified CWS algorithm \cite{zuo2019collector} and innovatively designs the CSI structure to improve the system reliability.
However, they struggle to manage geographical constraints effectively, making it challenging to avoid cable crossing issues.
Therefore, the infeasible solution obtained by the CWS algorithm needs refinements, and the final layout are depicted in Fig. \ref{91WT}(a) and (b).
In comparison, Cases 7 and 8 employ MP methods, with the key difference being the consideration of interconnection between OSSs. Their cable layout solutions are shown in Fig. \ref{91WT}(c) and (d). The costs corresponding to each structure are detailed in Table \ref{91WTplanning}. \par


\begin{table}[!htbp]
\begin{center}
\caption{Race Bank OWF ECS planning results}
\vspace{-1em}
\label{91WTplanning}
\setlength{\tabcolsep}{3.7mm}
\renewcommand{\arraystretch}{1.3}
\begin{tabular}{ccccc}
\hline\hline
Case No. & \begin{tabular}[c]{@{}c@{}} 5(\hspace{-0.08em}\cite{zuo2021two})\end{tabular} & \begin{tabular}[c]{@{}c@{}}6(\hspace{-0.08em}\cite{zuo2019collector})\end{tabular} & \begin{tabular}[c]{@{}c@{}}7\end{tabular} & \begin{tabular}[c]{@{}c@{}}8\end{tabular} \\
\hline
$C_{INV} (M$\$$)$ & 42.86 & 45.25 & 40.27 & 38.82 \\
$C_{O\&M} (M$\$$)$ & 21.03 & 22.20 & 19.76 & 19.04 \\
$C_{REL} (M$\$$)$ & 0.13 & 0.10 & 1.41 & 2.91 \\
Total cost $(M$\$$)$ & 64.02 & 67.54 & 61.44 & 60.77 \\
\hline\hline
\end{tabular}
\end{center}
\vspace{-1em}
\end{table}

Surprisingly, the least economical solution turns out to be Case 6, where the CSI structure has the highest investment and maintenance costs, yet its reliability is almost the same as the pure ring structure of Case 5. In the Race Bank OWF, the CSI structure does not achieve the anticipated benefits. This discrepancy arises because Ref. \cite{zuo2019collector} assumes that few geographical restrictions exist, allowing WTs to be uniformly distributed geographically. The superiority of the CSI is demonstrated through examples of OWFs that comply with this assumption. However, some OWFs such as Race Bank, may be subject to geographical limitations that prevent WTs from being uniformly distributed. In such cases, the CSI structure cannot be effectively implemented.
Since the layout produced by Case 5 is restricted to an isolated ring topology, it also leads to a very high cost. Excessive emphasis on reliability leads to a sub-optimal solution, with the total cost being about 3.25$M\$$ higher than that of Case 8.
The total costs for Cases 7 and 8 are relatively close, as both attain their respective optimal solutions through the MP method. However, the comparison of multiple costs reveals that by considering OSS coordination in Case 8, one can reduce cable length with a minor trade-off in reliability, thereby achieving the lowest total cost. This phenomenon can be interpreted from an optimization perspective: by removing structural restrictions and considering interconnections between OSSs, a gradual ``relaxation'' is achieved. As the feasible region expands, better solutions can be obtained.\par
\begin{figure}[!htbp]
\centering
\includegraphics[width=3.5in]{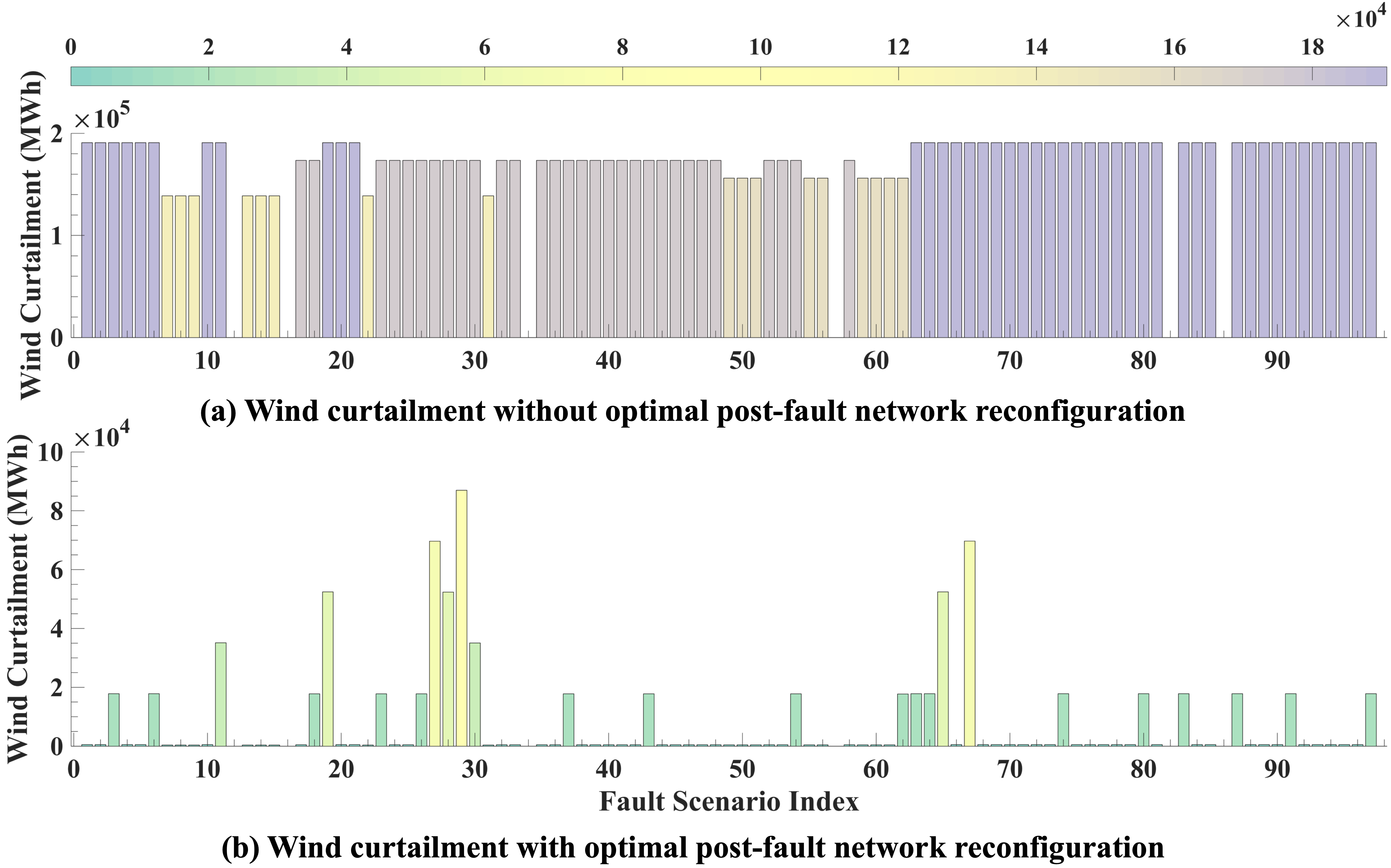}
\vspace{-2em}
\caption{Wind curtailment of contingency scenarios.}
\label{PFNR}
\end{figure}

The wind curtailment of Case 8 in various contingency scenarios, with and without considering post-fault network reconfiguration, is depicted in Fig. \ref{PFNR}. The scenario index aligns with the cable index in \cite{Ding2023}. By comparison, we can observe that the hybrid ECS structure and the implementation of optimal network reconfiguration greatly reduce the wind power curtailment and improve the system reliability effectively.\par
In summary, the comparative experiments demonstrate the superiority and scalability of the proposed method. Moreover, the OSS coordination and optimal network reconfiguration strategies in large OWFs contribute to a better balance between economy and reliability. \par

\vspace{-1em}
\subsection{Efficiency Validation of CPCI Algorithm}
\begin{figure}[!htbp]
\centering
\includegraphics[width=0.5\textwidth]{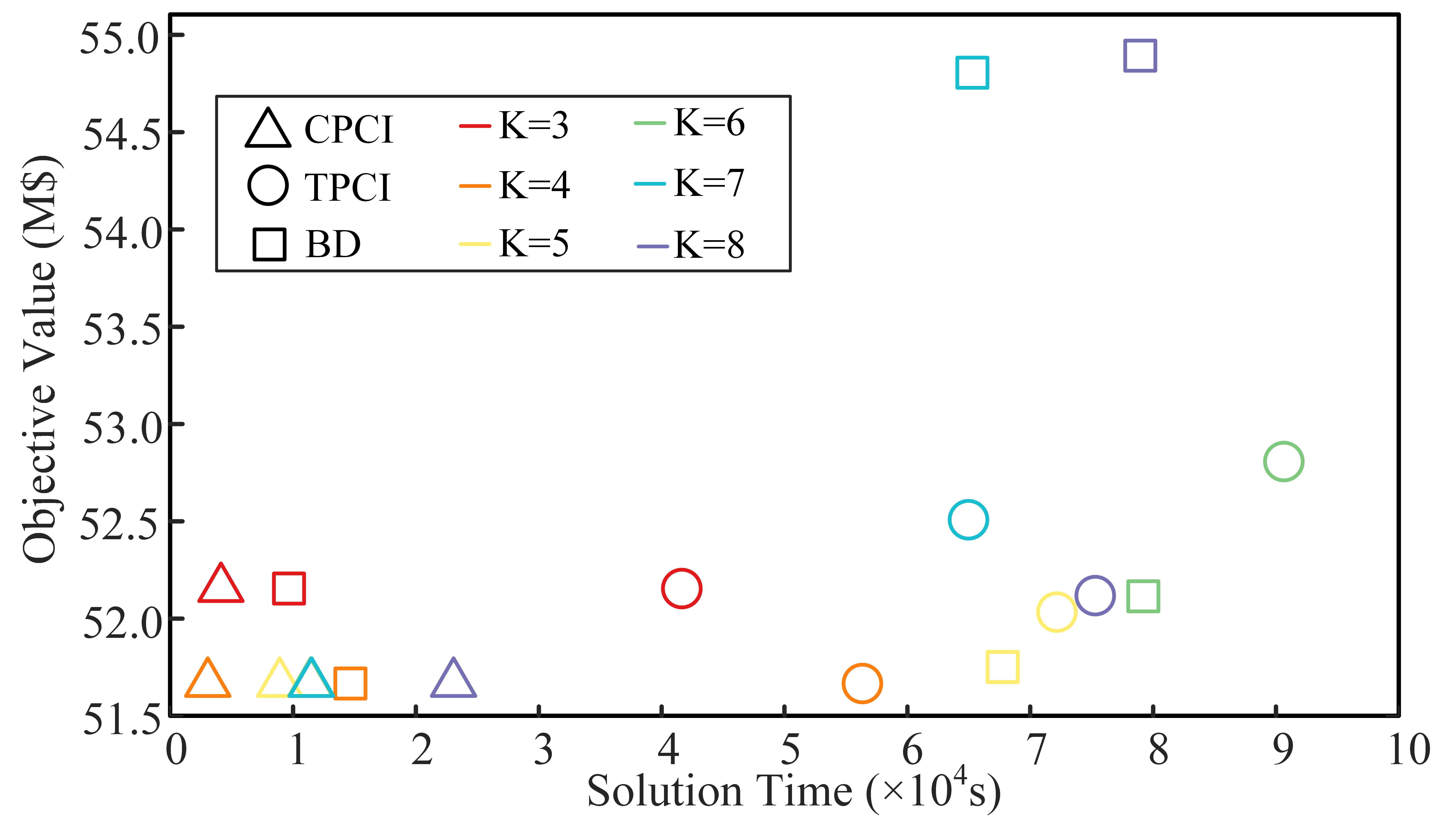}
\vspace{-2em}
\caption{Comparison of solution framework on Triton OWF. }
\label{SolutionFramework}
\end{figure}

To assess the advantages of the proposed CPCI algorithm, we compare it with TPCI\cite{lumbreras2012progressive} and BD algorithms. For each algorithm framework, the proposed model is solved for Triton OWF\cite{tritonknoll}, which comprises 90 WTs and 2 OSSs. The performances of frameworks are observed by varying K-values in the KNN algorithm. Since all three frameworks are exact solution frameworks, the optimality gap is set to 0. 
To balance efficiency, we cap the maximum number of iterations as 10 for each OWF partition in CPCI and TPCI. Additionally, each iteration is allotted a time frame of 2 hours. The results of this comparative study are shown in Fig. \ref{SolutionFramework}. \par
When the K-value is small, the results from CPCI, TPCI, and BD are consistent, indicating that CPCI operates as an exact algorithm. 
It can be found that CPCI consistently outperforms TPCI and BD in terms of both result quality and solution time. Initially, BD surpasses TPCI in solution efficiency. However, as the K-value increases, the problem complexity rises due to the growing number of candidate cables. Both TPCI and BD struggle to find the optimal solution within an acceptable timeframe. Notably, BD produces worse results than TPCI when the K-value is large. In contrast, CPCI consistently achieves the optimal solution rapidly, demonstrating its superiority over TPCI and BD. The robust performance of CPCI across different K-values underscores its efficacy.\par
The optimal solutions acquired through CPCI remain consistent when the K-value ranges from 4 to 8.
This validates the rationality of the divide-and-conquer approach, i.e., the optimal layouts do not incorporate excessively long cables. Generally, an increase in K-value makes the problem more complex and challenging to solve. After several tests, it is found that a K-value between 4 and 6 is most appropriate.\par

\begin{figure*}[!htbp]
\centering
\includegraphics[width=\textwidth]{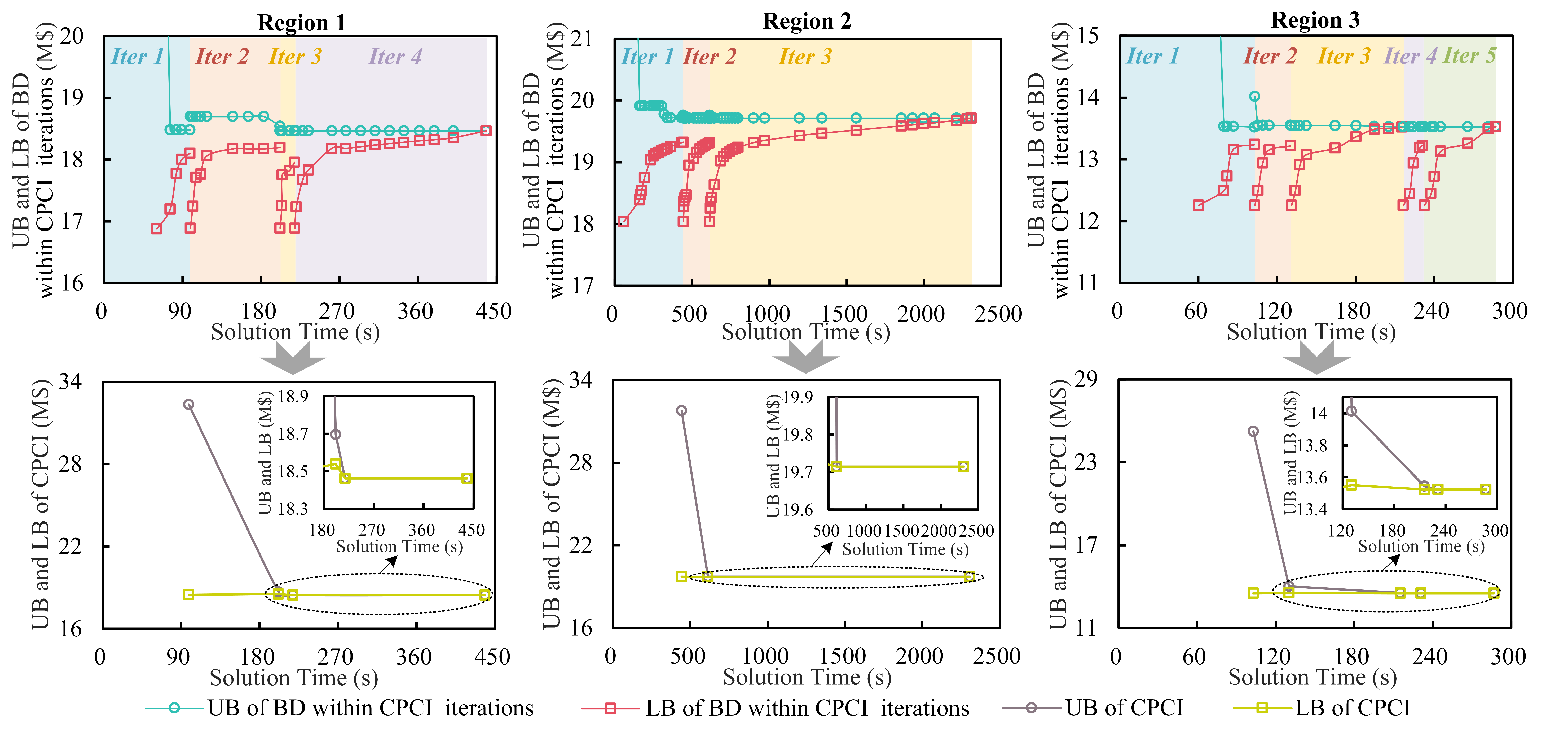}
\vspace{-2em}
\caption{Convergence trajectories of BD and CPCI algorithms. }
\label{Trajectories}
\vspace{-2em}
\end{figure*}


When the K-value is set to 4, the convergence trajectories of the BD and CPCI algorithms across regions are depicted in Fig. \ref{Trajectories}. The top three graphs illustrate the convergence trajectories of BD algorithm within CPCI iterations, while the bottom three graphs display the convergence trajectories of CPCI algorithm.
In the upper graphs, different CPCI iterations are marked by distinct background colors. 
The BD convergence trajectories show that in the initial CPCI iterations, the BD algorithm does not achieve convergence, with the upper bounds (UBs) and lower bounds (LBs) remaining unequal. Full convergence is attained only in the final CPCI iteration, which typically requires a longer solution time. This observation aligns with the proposed early termination and double-check mechanism, and confirms its efficacy in accelerating the solution process. 
The CPCI convergence trajectories indicate that there is no direct correlation between the number of iterations and the solution time. Notably, although Region 2 has the fewest iterations, it exhibits the longest solution time. The extended duration can be attributed to the increased complexity arising from the interconnection of OSSs in this region.\par
\begin{figure}[!htbp]
\centering
\includegraphics[width=0.5\textwidth]{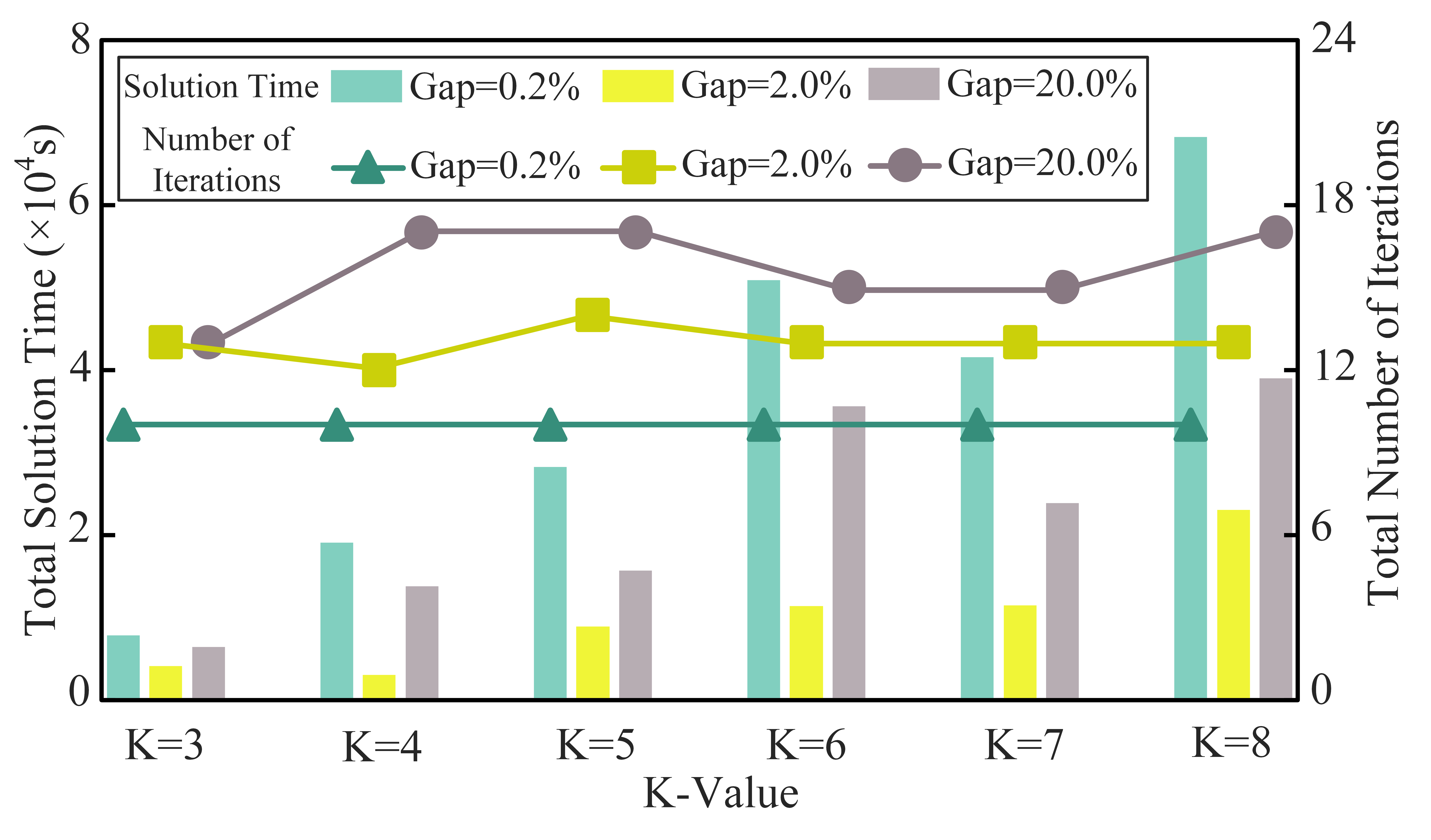}
\vspace{-2em}
\caption{Sensitivity analysis on CPCI gap tolerance $\epsilon_0$ and K-value. }
\label{Sensitivity}
\vspace{-1em}
\end{figure}

The solution performance of CPCI is significantly influenced by optimality gap tolerance $\epsilon_0$ in the iteration process, as detailed in Fig. \ref{Sensitivity}.
The sensitivity analysis of gap tolerance reveals a positive correlation between the gap and the number of iterations. 
However, setting a minor gap tolerance of 0.2\% results in protracted optimization times within each iteration, consequently extending the convergence duration of CPCI. On the flip side, adopting a substantially larger gap tolerance, like 20\%, yields lower-quality solutions within a single iteration. This situation causes the algorithm to incessantly fail the \textit{complete incorporation condition} checks and revert to further iterations. 
In comparison, opting for a moderate gap of 2.0\% allows a better trade-off between optimization time and result quality during the iteration process, thereby accelerating CPCI solving. The efficiency gains range from 36.2\% to 84.2\%.
\par

\section{Discussions}
The numerical results demonstrate the effectiveness and scalability of the proposed ECS planning model. In the original model, we made some assumptions to reduce its complexity. However, the assumptions can be relaxed to further consider cable type selection and simultaneous faults of multiple cables to make the model more applicable to real-world scenarios. This section provides detailed explanations on these aspects.

\vspace{-1em}

\subsection{Adaptability to Multi-Fault Scenarios}

It is noteworthy that multi-fault scenarios can be considered by expanding the contingency set in Assumption 2) and restructuring the ECS planning model. Now we define $u$ to represent different system operation scenarios, with the set of all fault scenarios denoted as $\mathcal{U}$. We define the set $\Psi_u$ to represent the set of cables that fail simultaneously in scenario $u$, given as $\Psi_u=\{(rs)_1, (rs)_2, \dots, (rs)_{|\Psi_u|}\}$. \par

In the original model, \( rs \in \Psi_C \) was used to represent single-fault operation scenarios of the system. Now, we replace the superscript \(rs\) with \(u\) to describe multi-fault operation scenarios of the system. Additionally, the original constraints should now hold for any multi-fault scenario \(u\in \mathcal{U}\).
Specifically, the fault impact identification constraint (12) and cable unavailability constraint (19) are rewritten as follows:

\begin{gather}
h_{k}^{f}+h_{rs}^{f}-1 \leq m_{k}^{rs}, \forall f \in \Psi_F, \forall k \in \Psi_N^{WT}, \forall rs \in \Psi_u
\\
m^u_k \geq m^{rs}_k, \forall rs \in \Psi_u, \forall u \in \mathcal{U}
\\
m^u_k \leq \sum_{rs \in \Psi_u} m^{rs}_k, \forall u \in \mathcal{U}
\end{gather}
\begin{equation}
s_{rs}^{u}=0, \forall rs \in \Psi_u, \forall u \in \mathcal{U}
\end{equation}

The modified ECS planning model considering multi-fault scenarios is formulated as:
$$
\begin{gathered}
\quad \min \text{(1)} \\
s.t. \text { (2)-(11), (13)-(18), (20)-(27), (37)-(40)}.
\end{gathered}
$$\par
\vspace{-1em}

The restructured model takes the form of an MILP, and retains the \textit{contingency structure} defined in \textbf{Definition 1}. Clearly, the CPCI algorithm is still applicable to the modified model and similarly converges when the \textit{complete incorporation condition} is satisfied. The only difference is that scenarios involving simultaneous faults of multiple cables should also be considered to update the subset $\Tilde{\Upsilon}$ during the iterative process.

\vspace{-1em}

\subsection{Adaptability to Cable Type Selection}

In order to take cable type selection into consideration, it is necessary to introduce the binary variable $l_{ij}^v$, which equals 1 if cable $ij$ with type $v$ is invested for installation. Here, $v$ represents the cable type, with $v \in V$.
The cable investment cost in the objective function can be rewritten as:

\begin{equation}
C_{INV}= \sum_{v \in V} \sum_{i j \in \Psi_C} C_{cab}^v a_{ij} l_{i j}^v,
\end{equation}
where $C_{cab}^v$ represents the cost of purchasing and installing one unit length of submarine cable with type $v$.\par

Due to the different capacities of various cable types, constraint (9) is rewritten as follows:
\begin{equation}
-\sum_{v \in V} P^{C,v} l_{ij}^v \leq P_{i j}^{rs} \leq \sum_{v \in V} P^{C,v} l_{ij}^v, \forall i j \in \Psi_C,
\end{equation}
where $P^{C,v}$ denotes the capacity of the cable with type $v$.
\par

Constraint (43) needs to be added to ensure that at most one type of cable is installed at the same location.
\begin{equation}
\sum_{v \in V} l_{ij}^v = l_{ij}, \forall ij \in \Psi_C
\end{equation}

The modified ECS planning model considering cable type selection is formulated as:
$$
\begin{gathered}
\quad \min \text{(1)} \\
s.t. \text { (3)-(8), (10)-(27), (41)-(43)}.
\end{gathered}
$$

\par

In the updated stochastic programming model, the first-stage variables are $\{l_{ij}, l_{ij}^v\}$, and the second-stage variables are $\{s_{ij}^{NO/rs}, h_{k}^f, h_{ij}^f, m_{k}^{NO/rs}, n_{k}^{NO/rs}, P_{f}^{rs}, P_{ij}^{rs}, P_{k}^{rs}, \theta_{i}^{rs}\}$.
The CPCI algorithm remains applicable. However, due to the consideration of cable type selection, the model inevitably becomes more complex, potentially leading to significantly longer solution times for each iteration. Therefore, more powerful algorithms need to be developed in future research to accelerate the solution process.

\par
\vspace{1em}
\section{Conclusion}
This paper proposes a novel ECS cable layout planning method based on two-stage stochastic programming.
Unlike conventional approaches, this method does not predefine the network topology.
Moreover, the optimal network reconfiguration strategies under various contingencies are included.
To deal with the dimension curse of large-scale ECS planning, a divide-and-conquer approach and CPCI algorithm are developed. The conclusions are summarized as follows:
\begin{itemize}
\item[1)] Generally, the optimal ECS structure deviates from typical radial or ring patterns. The optimal layout can achieve cost savings of around 20\% compared to radial and ring layouts in the 30-WT case. 
This finding supports the necessity of fully optimizing the ECS structure rather than restricting the design to predefined configurations.
\item[2)] The MP methods, with the flexibility to impose geographical constraints and avoid infeasible outputs, produce more balanced results than heuristic methods. Optimal post-fault network reconfiguration strategies greatly reduce wind power curtailment, playing a crucial role in reliability enhancement. With the consideration of OSS coordination, a more favorable trade-off between cost-effectiveness and reliability can be achieved.
\item[3)] Tests validate the efficacy of the divide-and-conquer approach and suggest suitable K-values to balance optimality and solution efficiency. The proposed CPCI algorithm achieves optimal solutions rapidly and exhibits superior performance compared to TPCI and BD algorithms. Sensitivity analysis indicates that a moderate CPCI gap can improve efficiency, with gains ranging from 36.2\% to 84.2\%.
\end{itemize}
\par
Despite its good performance, this work still requires further investigation. This paper limits its focus to a single cable type to simplify the model. While it demonstrates the feasibility of adapting to multiple cable types, the complexity of more advanced models poses significant computational challenges. Future research could focus on developing more efficient solution techniques, e.g., by incorporating machine learning into the branch and bound process of the modified model, as well as improving the strategy for generating cuts.
Additionally, employing stochastic programming method presupposes that wind speeds follow a known distribution. However, if the exact probability distribution of wind speed is difficult to acquire, stochastic programming may not be appropriate. This limitation could potentially be overcome by using robust optimization or distributionally robust optimization methods in future research.
Moreover, the positions of WTs and OSSs are preset in this study. However, the joint optimization of micro-siting and cable layout planning might produce enhanced results from a more comprehensive perspective. Future research could investigate the interaction between them with a multi-level framework.



%

\vspace{-1em}
\appendix[Proof of Remark 5]
According to \textbf{Remark 3}, we know that the optimal value $\Tilde{z}^*$ of $\mathcal{\Tilde{P}}_{\Tilde{\Upsilon},\Omega}(x,y^{\upsilon,\omega})$ is necessarily lower than or equal to the optimal value $z^*$ of $\mathcal{P}_{\Upsilon,\Omega}(x,y^{\upsilon,\omega})$. In other words, $\Tilde{z}^*$ provides a lower bound for $z^*$.\begin{equation}
\mathcal{\Tilde{P}}_{\Tilde{\Upsilon},\Omega}(x^*,y^{\upsilon,\omega *})=\Tilde{z}^* \leq z^*
\end{equation}
When the \textit{complete incorporation condition} is satisfied, the difference between $\mathcal{\Tilde{P}}_{\Tilde{\Upsilon},\Omega}(x,y^{\upsilon,\omega})$ and $\mathcal{P}_{\Upsilon,\Omega}(x,y^{\upsilon,\omega})$ lies in the consideration of faulty operational state scenarios related to non-installed cables, which do not have any impact. Therefore, ($x^*$, $y^{\upsilon,\omega *}$) is a feasible solution to $\mathcal{P}_{\Upsilon,\Omega}(x,y^{\upsilon,\omega})$, and substituting it yields an upper bound for the problem.
\begin{equation}
z^* \leq {z^*}^\prime=\mathcal{P}_{\Upsilon,\Omega}(x^*,y^{\upsilon,\omega *})
\end{equation}
According to \textbf{Remark 4}, the following equations hold.
\begin{equation}
\begin{aligned}
&\mathcal{\Tilde{Q}}_{\Tilde{\Upsilon},\Omega}(x^*)=\mathcal{Q}_{\Upsilon,\Omega}(x^*)\\
& =\min _{y^{\upsilon,\omega}}\sum_{\substack{\upsilon \in \Upsilon^* \\ \upsilon\neq \upsilon_0}} \xi^\upsilon \sum_{\omega \in \Omega}\delta^\omega  d^Ty^{\upsilon,\omega}
+\small(1-\sum_{\substack{\upsilon \in \Upsilon^* \\ \upsilon\neq \upsilon_0}} \xi^\upsilon\small) \sum_{\omega \in \Omega} \delta^\omega  d^Ty^{\upsilon_0,\omega}
\end{aligned}
\end{equation}
\begin{equation}
\begin{aligned}
& \Tilde{z}^*=\mathcal{\Tilde{P}}_{\Tilde{\Upsilon},\Omega}(x^*,y^{\upsilon,\omega *})=c^Tx^*+\mathcal{\Tilde{Q}}_{\Tilde{\Upsilon},\Omega}(x^*) \\
& =c^Tx^*+\mathcal{Q}_{\Upsilon,\Omega}(x^*)=\mathcal{P}_{\Upsilon,\Omega}(x^*,y^{\upsilon,\omega *})={z^*}^\prime
\end{aligned}
\end{equation}
The equality of the upper and lower bounds indicates that ($x^*, y^{\upsilon, \omega *}$) is the optimal solution to $\mathcal{P}_{\Upsilon,\Omega}(x, y^{\upsilon, \omega})$ as well, which completes the proof. 
$\hfill\blacksquare$ 
\vspace{-1em}



\ifCLASSOPTIONcaptionsoff
  \newpage
\fi


\normalem
\bibliographystyle{IEEEtran}
\bibliography{references.bib}

\end{document}